\documentclass{elsarticle}

\pdfoutput=1

\usepackage{graphicx}
\usepackage{amsmath}
\usepackage{amsfonts}
\usepackage{amssymb}
\usepackage{upgreek}
\usepackage{mathtools}
\usepackage{url}
\usepackage{tikz}

\usepackage{listings}
\definecolor{cmtgray}{gray}{0.45}
\definecolor{cmtmgray}{gray}{0.20}
\lstdefinestyle{customc}{
  belowcaptionskip=.3\baselineskip,
  escapeinside={/*@}{@*/},
  breaklines=true,
  frame=tB,
  xleftmargin=\parindent,
  language=C++,
  showstringspaces=false,
  basicstyle=\ttfamily,
  keywordstyle=\bfseries\color{green!40!black}\textbf,
  commentstyle=\color{cmtgray}\ttfamily,
  identifierstyle=\color{black!90!white},
  stringstyle=\color{black},
  captionpos=b,
  linewidth=\textwidth,
 morekeywords={}
}
\usepackage{color}

\newcommand{\bigoo}{\mathcal{O}}

\begin{document}

\begin{frontmatter}

\title{Task parallel implementation of a solver for electromagnetic scattering problems\tnoteref{mytitlenote}}
\tnotetext[mytitlenote]{This article is based upon work from COST Action IC1406 cHiPSET, supported by COST (European
Cooperation in Science and Technology). The authors are ordered with respect to affiliation first, and role in the project secondly.}

\author[uu]{Afshin Zafari}
\ead{afshin.zafari@it.uu.se}

\author[uu]{Elisabeth Larsson\corref{cor}}
\ead{elisabeth.larsson@it.uu.se}
\cortext[cor]{Corresponding author}

\author[ismb]{Marco Righero}
\ead{righero@ismb.it}

\author[ismb]{M. Alessandro Francavilla}
\ead{francavilla@ismb.it}

\author[ismb]{Giorgio Giordanengo}
\ead{giordanengo@ismb.it}

\author[polito]{Francesca Vipiana}
\ead{francesca.vipiana@polito.it}

\author[polito]{Giuseppe Vecchi}
\ead{giuseppe.vecchi@polito.it}

\address[uu]{Dept.\ of Information Technology, Uppsala University, Box 337, SE-751~05 Uppsala, Sweden }
\address[ismb]{Antenna and EMC Lab (LACE),
Istituto Superiore Mario Boella, Torino 10138, Italy}
\address[polito]{Antenna and EMC Lab (LACE),
Politecnico di Torino, Torino 10129, Italy}



\begin{abstract}
Electromagnetic computations, where the wavelength is small in relation to the geometry of interest, become computationally demanding. In order to manage computations for realistic problems like electromagnetic scattering from aircraft, the use of parallel computing is essential. In this paper, we describe how a solver based on a hierarchical nested equivalent source approximation can be implemented in parallel using a task based programming model. We show that the effort for moving from the serial implementation to a parallel implementation is modest due to the task based programming paradigm, and that the performance achieved on a multicore system is excellent provided that the task size, depending on the method parameters, is large enough. 
\end{abstract}


\end{frontmatter}

\section{Introduction}
Simulations of electromagnetic fields~\cite{1979_Mautz_Harrington_Electromagnetic} are performed in industry in many different applications. One of the most well known is antenna design for aircraft. But electromagnetic behavior is important, e.g., also for other types of vehicles, for satellites, and for medical equipment.
\begin{figure}[!htb]
\centering
\includegraphics[width=0.7\textwidth]{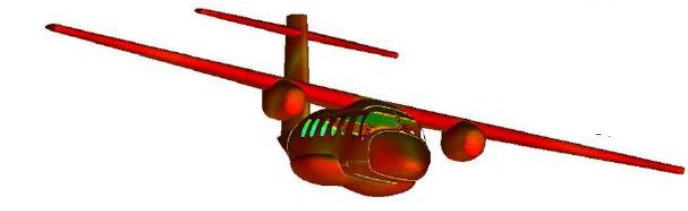}
\caption{Surface currents on an aircraft model from a boundary element simulation with around 2 million unknowns.}\label{fig:1}
\end{figure}
A common way to reduce the cost of the numerical simulation is to assume time-harmonic solutions, and to reformulate the Maxwell equations describing the electromagnetic waves in terms of surface currents. That is, the resulting numerical problem is time-independent and is solved on the surface of the body being studied, see Figure~\ref{fig:1} for an example of a realistic aircraft surface model.

The size $N$ of the discretized problem, which for a boundary element discretization takes the form of a system of equations with a dense coefficient matrix, can still be very large, on the order of millions of unknowns going up to billions, and this size increases with the wave frequency. If an iterative solution method is applied to the full (dense) matrix, the cost for each matrix-vector multiplication is $\bigoo(N^2)$, and direct storage of the matrix also requires memory resources of $\bigoo(N^2)$. Different (approximate) factorizations of the matrices, that can reduce the costs to $\mathcal{O}(N\log N)$ or even $\mathcal{O}(N)$, have been proposed in the literature such as the MultiLevel Fast Multipole Algorithm (MLFMA), see, e.g.,~\cite{1997_SongETAL_Multilevel}; FFT-based factorization, see, e.g.,~\cite{2005_SeoLee_AFast,2010_Vipiana_ETAL_EFIEModeling}; factorizations based on the Adaptive Cross Approximation (ACA), see, e.g.,~\cite{2005_ZhaoETAL_TheAdaptive}; or based on H2 matrices as the Nested Equivalent Source Approximation (NESA)~\cite{2014_Li_ETAL_Nested, 2014_Li_ETAL_ADoubly, 2015_Li_ETAL_Wideband}. 
 
All these approximations can be seen as decomposing the original dense matrix into a sparse matrix accounting for near field interactions, and a hierarchical matrix structure with low storage requirements accounting for far field interactions.

This presents some problems for the parallel implementation required for large scale problems. The resulting algorithm is typically hierarchical with interactions both in the vertical direction between parents and children in a tree structure, and horizontally between the different branches at each level of the tree. Furthermore, the tree can be unbalanced in different ways due to the geometry of the underlying structure. The article~\cite{Darve11} provides an overview of the challenges inherent in the implementation of the fast multipole method (FMM) on modern computer architectures. 

There is a rich literature on parallel implementation of FMM-like algorithms. The classical approach, targeting distributed memory systems, is to partition the tree data structure over the computational nodes, and use an MPI-based parallelization~\cite{Kurzak05}. The resulting performance is typically a bit better for volume formulations then for boundary formulations, since the computational density is higher in the former case. A particular issue for the MLFMA formulation of electromagnetic scattering problems is that the work per element (group) in the tree data structure increases with the level, and additional partitioning strategies are needed for the coarser part of the structure~\cite{Velam05,GuEr13,Benson14}.
 
The ongoing trend in cluster hardware is an increasing number of cores per computational node. When scaling to large numbers of cores, it is hard to fully exploit the computational resources using a pure MPI implementation~\cite{Zafari17}. As is pointed out in~\cite{Lashuk12}, a hybrid parallelization with MPI at the distributed level and threads within the computational nodes is more likely to perform well. That is, a need for efficient shared memory parallelizations of hierarchical algorithms to be used in combination with the distributed MPI level arises.

The emerging method of choice for implementing complex algorithms on multicore architectures is dependency-aware task-based parallel programming, which is available, e.g, through the StarPU~\cite{ATNW11}, OmpSs~\cite{PeBaLa08}, and SuperGlue~\cite{Tillenius15} frameworks, but also in OpenMP, since version 4.0. Starting with~\cite{Bordage12}, where StarPU is used for a task parallel implementation of an FMM algorithm, several authors have taken an interest in the problem. In~\cite{HEGH14}, SuperGlue is used for a multicore CPU+GPU implementation of an adaptive FMM. The Quark~\cite{icl:609} run-time system is used for developing an FMM solver in~\cite{Ltaief14}. Since tasks were introduced in OpenMP, a recurring question is if the OpenMP implementations can reach the same performance as the specific run-times discussed above. An early OpenMP task FMM implementation is found in~\cite{Agullo14}. This was before the depend clause was introduced, allowing dependencies between sibling tasks. OpenMP, Cilk and other models are compared for FMM in~\cite{Zhang14}, OpenMP and Klang/StarPU are compared in~\cite{Agullo17}, and different OpenMP implementations and task parallel run-times are compared with a special focus on locking and synchronization in~\cite{Atkinson17}. A common conclusion from these comparisons is that the commutative clause provided by most task parallel run-time systems is quite important for performance, and that this would be a useful upgrade of OpenMP tasks for the future.

An alternative track is to develop special purpose software components for the class of FMM-like algorithms, see, e.g., PetFMM~\cite{Cruz11} and Tapas~\cite{Fukuda17}.



In this paper, we use the SuperGlue~\cite{Tillenius15} framework, and implement a model problem that is similar to the NESA algorithm for electromagnetic scattering. We investigate how the task size influences the performance, and show how the task parallel model allows for a mixing of the computational phases that scales well even when the individual phases suffer form scalability issues. We also compare the results with an OpenMP task implementation.

The paper is organized as follows: In Section~\ref{sec:ienesa}, we briefly recap the basics of the integral equation formulation of electromagnetic scattering and radiation problems, and discuss some efficient algorithms for solving the discretized problem. In Section~\ref{sec:model}, we describe the simplified two-dimensional problem that is used for evaluating the parallelization approach. Section~\ref{sec:taskparallel} provides a general introduction to task based parallel programming, while Section~\ref{sec:impl} is focused on the implementation of the specific algorithm investigated here. In Section~\ref{sec:perf}, the parallel performance of the new implementation is evaluated. In Section~\ref{sec:omp}, the comparison with OpenMP is discussed, and finally in Section~\ref{sec:sum}, the results are summarized.

\section{Integral equation formulation and the nested equivalent source approximation}
\label{sec:ienesa}

Let $\Omega$ be a volume, whose boundary is $\partial \Omega$, surrounded by a homogeneous medium with wavenumber $k$ and intrinsic impedance $\eta$. The electric field $\boldsymbol{e}^s$ at $\boldsymbol{r}$, a point in the exterior medium, radiated by a current $\boldsymbol{j}$ on the surface $\partial \Omega$ is given by
\begin{equation}
\label{eq:stratton}
\boldsymbol{e}^s(\boldsymbol{j})(\boldsymbol{r}) = -\mathrm{i}\eta k \left(\int_{\partial \Omega}g(\boldsymbol{r}, \boldsymbol{r}')\boldsymbol{j}(\boldsymbol{r}')\mathrm{d}\boldsymbol{r}' 
+ \frac {1}{k^2}\nabla \int_{\partial \Omega} g(\boldsymbol{r}, \boldsymbol{r}') \div \boldsymbol{j}(\boldsymbol{r}')\mathrm{d}\boldsymbol{r}'\right), 
\end{equation}
where $g(\boldsymbol{r}, \boldsymbol{r}') = \exp(-\mathrm{i} k \|\boldsymbol{r} - \boldsymbol{r}'\|)/(4\uppi \|\boldsymbol{r} - \boldsymbol{r}'\|)$ is called the Green's function of the exterior medium.
If an external electric field $\boldsymbol{e}^i$ impinges on the surface $\partial \Omega$, we can determine the resulting current on $\partial \Omega$ enforcing appropriate boundary conditions on $\partial \Omega$ and, in turn, use this current to evaluate the radiated field in any position. When considering a metallic object, from a numerical viewpoint, we expand the unknown current in terms of an appropriate basis,
$
\boldsymbol{j}(\boldsymbol{r}) = \sum_{n = 1}^N j_n \boldsymbol{\phi}_n(\boldsymbol{r}),
$
and enforce a null-field condition in the weak sense, namely we impose
\begin{equation}
\int_{\partial \Omega}\boldsymbol{e}^s(\boldsymbol{j})(\boldsymbol{r}) \cdot \boldsymbol{\phi}_m(\boldsymbol{r}) = - \int_{\partial \Omega}\boldsymbol{e}^i(\boldsymbol{r}) \cdot \boldsymbol{\phi}_m(\boldsymbol{r}), \quad \forall m = 1, \ldots, N.
\end{equation}
This results in a linear system 
$
\left[Z\right] \left[j\right] = - \left[e^i\right],
$
to be solved to determine coefficients $j_n$. The entries of the matrix $\left[Z\right]$ and the vector $\left[e^i\right]$ are given by
\begin{equation}
Z_{m n}   = \int_{\partial \Omega}\boldsymbol{e}^s(\boldsymbol{\phi}_n)(\boldsymbol{r}) \cdot \boldsymbol{\phi}_m(\boldsymbol{r})\mathrm{d} \boldsymbol{r} ,\quad
e^i_{m}   = -\int_{\partial \Omega}\boldsymbol{e}^i(\boldsymbol{r}) \cdot \boldsymbol{\phi}_m(\boldsymbol{r})\mathrm{d} \boldsymbol{r}.
\end{equation}
When considering large structures, the resulting linear system is usually solved with an iterative solver, such as BiCGStab or GMRES.

The formulation sketched here  is suitable for both scattering and radiation problems. However, it poses three main difficulties, related to the fact that matrix entries are given as convolution integral with a non-local kernel $g(\boldsymbol{r}, \boldsymbol{r}')$: Forming the matrix $\left[Z\right]$, storing the matrix, and performing the MVPs required by any iterative solver.

As the kernel $g(\boldsymbol{r}, \boldsymbol{r}')$ is smooth (when $\boldsymbol{r} \neq \boldsymbol{r}'$) and decreases as $1/\|\boldsymbol{r} - \boldsymbol{r}'\|$, portions of the matrix corresponding to well separated parts of the surface $\partial \Omega$ can be approximated with low-complexity factorizations. 

Without going into the details explained in \cite{2014_Li_ETAL_Nested}, we here sketch the basic idea, for ease of reference. The structure is hierarchically partitioned using an oct-tree. Portions of the matrix $[Z]$ corresponding to interaction between basis functions in near leaf blocks are computed and stored directly, in a sparse matrix $\left[Z\right]^{\mathrm{near}}$. Portions of the matrix corresponding to interaction between basis functions in far blocks are computed and stored in an approximate structured way.

We denote with $\alpha$ and $\beta$ two far groups at the leaf level $L$, and with $P(\alpha)$ the parent group of the group $\alpha$. Let $\left[Z\right]_{\alpha, \beta}$ denote the block of the matrix $[Z]$ corresponding to the interactions between basis functions in $\alpha$ end $\beta$, with size $T\times S$. In low-medium frequency regimes, it is rank deficient \cite{1997_Kapur_Long_IES3, 2000_Bebendorf_Approximation} and we can approximate it as
\begin{equation}
\label{eq:nesa:udv}
\left[Z\right]_{\alpha, \beta} = \left[U\right]_{\alpha}\left[D\right]_{\alpha, \beta}\left[V\right]_{\beta}
\end{equation}
where $\left[U\right]_{\alpha}$, $\left[D\right]_{\alpha, \beta}$, and $\left[V\right]_{\beta}$ have sizes $T\times R$, $R\times R$, and $R\times S$, respectively, with $R\ll (T, S)$.

Applying the same idea to a child and its parent group, we see that we can approximate $\left[D\right]_{\alpha, \beta}$ as
\begin{equation}
\label{eq:nesa:bdc}
\left[D\right]_{\alpha, \beta} = 
\left[B\right]_{\alpha, P(\alpha)}
\left[D\right]_{P(\alpha) P(\beta)}
\left[C\right]_{P(\beta), \beta},
\end{equation}
so that we have the 1-level approximation
\begin{equation}
\label{eq:nesa:1level}
\left[Z\right]_{\alpha, \beta} = 
\left[U\right]_{\alpha}
\left[B\right]_{\alpha, P(\alpha)}
\left[D\right]_{P(\alpha) P(\beta)}
\left[C\right]_{P(\beta), \beta}
\left[V\right]_{\beta}.
\end{equation}
%
In general, if we move along the family tree induced by the oct-tree division for $\bar{\ell}$ levels, we have 
\begin{equation}
\label{eq:nesa:Llevel}
\begin{split}
\left[Z\right]_{\alpha, \beta}  =&
\left[U\right]_{\alpha}
\underbrace{\left[B\right]_{\alpha, P(\alpha)}
	\cdots
	\left[B\right]_{P^{\bar{\ell}-1}(\alpha), P^{\bar{\ell}}(\alpha)}}_{\mathrm{ascending}}
\left[D\right]_{P^{\bar{\ell}}(\alpha) P^{\bar{\ell}}(\beta)}\\
& \underbrace{\left[C\right]_{P^{\bar{\ell}}(\beta), P^{\bar{\ell}-1}(\beta)}
	\cdots
	\left[C\right]_{P(\beta), \beta}}_{\mathrm{descending}}
\left[V\right]_{\beta}.
\end{split}
\end{equation}
Matrices $[U], [D], [V], [B], [C]$ are built enforcing equivalence conditions on fictitious boundaries enclosing the blocks. Their construction is based on standard linear algebra matrix operations and has cost independent of $N$. From the way they are constructed, we interpret $[U]$ as \emph{receiving matrix}, $[D]$ as \emph{translation matrix}, $[V]$ as \emph{radiation matrix}, and $[B]$ and $[C]$ as \emph{transfer matrices}. 

If matrix $[Z]_{\alpha, \beta}$ collects the field due to currents in leaf block $\beta$, tested on functions in leaf block $\alpha$, we ascend the three up to level $\ell_0$, so that blocks $P^{L-\ell0}(\alpha)$ and $P^{L-\ell0}(\beta)$ are still not touching, and approximate $[Z]_{\alpha, \beta}$  using Equation~\ref{eq:nesa:Llevel} with $\bar{\ell} = L - \ell_0$.

When dealing with 2D structures, the algorithm can be used with little modifications, due to its purely algebraic nature: The Green's function in Equation~\ref{eq:stratton} becomes $g(\boldsymbol{r}, \boldsymbol{r}') = H^{(2)}_0(k\left| \boldsymbol{r} - \boldsymbol{r}'\right|)$, with $H^{(2)}_0$ being the Hankel function of the second kind, order zero, and surface integrals become line integrals. 

\section{The model problem}\label{sec:model}
To test the hypothesis that a task-based parallelization is suitable for electromagnetic scattering problems, we implement a simplified two-dimensional problem for a first evaluation, avoiding the complications of implementing the boundary element method, while focusing on the interaction structures when using the hierarchical NESA representation of a matrix.

The simplified problem consists of computing the two-dimensional electric potential $\phi(\boldsymbol{x})$ generated by charges (sources) located at the points $\boldsymbol{x}_i$, $i=1,\ldots,N$ with charges $q(\boldsymbol{x}_i)$, and evaluating it at the source locations. The dense matrix version of the problem takes the form
\begin{equation}
[\phi] = [K][q],
\label{eq:pot}
\end{equation}
where the matrix elements are given by $k_{ij}=K(\boldsymbol{x}_i,\boldsymbol{x}_j)=-\log(|\boldsymbol{x}_i-\boldsymbol{x}_j|)$, $i\neq j$, and $k_{ii}=0$, $i,j=1,\ldots,N$.
%

\begin{figure}[!htb]
\includegraphics[width=0.5\textwidth]{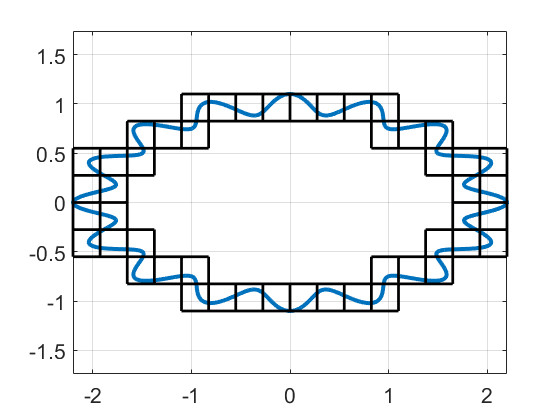}%
\includegraphics[width=0.5\textwidth]{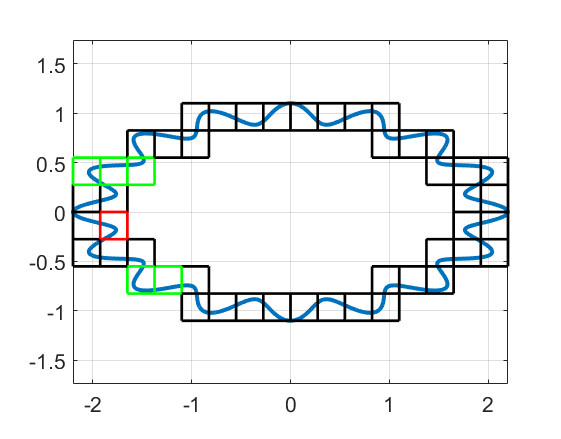}\\
\includegraphics[width=0.5\textwidth]{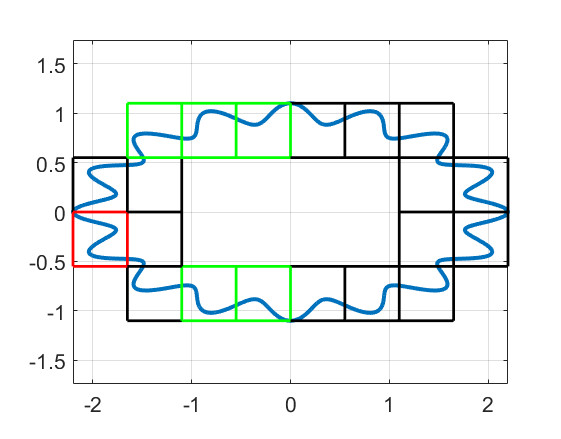}%
\includegraphics[width=0.5\textwidth]{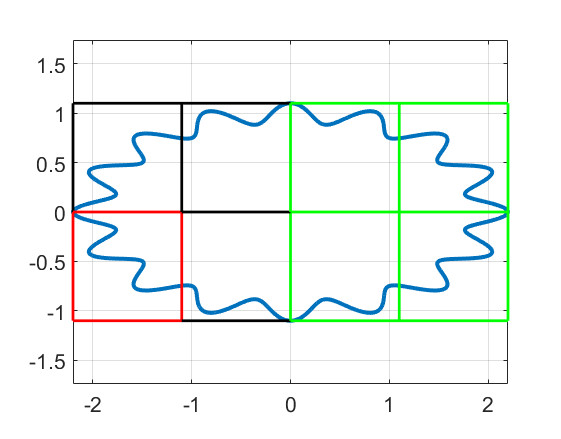}
\caption{The source points (charges) are located on the blue curve, and the potential is also evaluated at these points. Groups at three different levels are shown. The far field computation contributing to the smallest red group are also shown at each level. The interaction with the whole right half of the domain (marked with green color) is handled at the coarsest level, then additional groups are taken care of at each finer level until only the near field groups remain.}
\label{fig:boxes}
\end{figure}

In the model problem, the charges are located on a one-dimensional structure within the two-dimensional space. The particular curve that we have used as an example for our source locations is shown in Figure~\ref{fig:boxes}. To construct the NESA representation of the matrix, we start by constructing the hierarchical tree structure that should cover the domain of the sources. The tree is constructed with levels $\ell=\ell_0,\ldots,L$. The coarsest level $\ell_0$ is chosen such that we have 4 groups along the longest dimension of the domain, see Figure~\ref{fig:boxes}. The depth of the tree is determined by the method parameter $P$. The groups are subdivided as long as the average number of source points in the finest level groups does not fall below $P$. The example in Figure~\ref{fig:boxes} only shows three levels, but in the examples with $N=100\,000$ sources we are using in the numerical experiments, there are 8 active levels. 

\begin{figure}[!htb]
\centering
\includegraphics[width=0.4\textwidth]{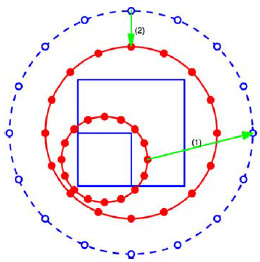}
\caption{A parent group and one of its children are illustrated. The points on the red circles are where the auxiliary sources are located, and the points on the blue circle are where the potentials are matched.}\label{fig:equiv}
\end{figure}

Figure~\ref{fig:boxes} also shows how the computations are divided into near and far field interactions. The near field interactions are computed at the finest level between groups that are close neighbors (left, right, top, bottom, diagonal) and within each group. The near field interaction between groups $\alpha$ and $\beta$ constitute reduced versions of the global problem~\eqref{eq:pot}
\begin{equation}
[\phi]_\alpha = [\phi]_\alpha + [K]_{\alpha,\beta}[q]_\beta.
\label{eq:algnf}
\end{equation}
Far field interactions are computed at each level. The groups that interact in the far field computations are the ones that are not near neighbors at the current level, and whose parents were not part of the interaction at the previous level. In this way, only a limited number of boxes are involved in the far field interaction at each level. If the interactions were computed directly, there would be no computational savings. This is where NESA comes in. Simply put, we compute equivalent charges in each group at each level. We then compute the fields created by the interaction of these charges at the same points, and then transfer the corresponding field to the actual source points. Figure~\ref{fig:equiv} shows the location of the equivalent source points in a parent and child group together with the test surface where the fields are matched in order to form the low rank approximation. The number of equivalent charges $Q$ is the same  in each group, which is why we can save significantly in the far-field computation.

We will not go into all details here, instead we refer to~\cite{2014_Li_ETAL_Nested}, but we will describe each step in the algorithm in a way that helps the later discussion of the parallel implementation. The far field interaction can be divided into 5 different stages:

\begin{description}
\item[Radiation:] For each finest level group $\beta$, the equivalent sources $s$ at the auxiliary points (see Figure~\ref{fig:equiv}) are computed from the actual sources
\begin{equation}
[s]_\beta = [V]_\beta[q]_\beta.
\label{eq:algrad}
\end{equation}

\item[Source transfer:] Next, equivalent sources are computed for every level of the tree. Each child group $\beta_\ell$ at each level $\ell$ transfers its charge to its parent group $P(\beta_\ell)$
\begin{equation}
[s]_{P(\beta_\ell)} = [s]_{P(\beta_\ell)} + [C]_{P(\beta_\ell),\beta_\ell}[s]_{\beta_\ell},\quad \ell_0+1\leq\ell\leq L.
\label{eq:algsou}
\end{equation}

\item[Translation:] For each level $\ell$ and each (observation) group $\alpha_\ell$ at that level, the contribution to the potential generated by the groups $\beta_\ell$ in the far-field interaction list at the same level is computed
\begin{equation}
[o]_{\alpha_\ell} = [o]_{\alpha_\ell} + [D]_{\alpha_\ell,\beta_\ell}[s]_{\beta_\ell}, \quad \ell_0\leq\ell\leq L.
\label{eq:algtrans}
\end{equation}

\item[Potential transfer:] The potential contribution at each parent group $P(\alpha_\ell)$ is transferred and added to its child group's potentials. 
\begin{equation}
[o]_{\alpha_\ell} = [o]_{\alpha_\ell} + [B]_{\alpha_\ell,P(\alpha_\ell)}[o]_{P(\alpha_\ell)} ,\quad \ell_0+1\leq\ell\leq L.
\label{eq:algpot}
\end{equation}


\item[Reception:] Finally, the potential at the auxiliary points of each finest level group $\alpha$ is transferred to the actual observation points.
\begin{equation}
[\phi]_\alpha = [\phi]_\alpha + [U]_\alpha[o]_\alpha.
\label{eq:algrec}
\end{equation}
\end{description}








\section{Task parallel programming}
\label{sec:taskparallel}
One of the key features of task parallel programming is that it makes it relatively easy for the programmer to produce a parallel application code that performs well. However, it is still important for the programmer to know how to write a task parallel program and how various aspects of the algorithm are likely to impact performance. 

As an example, we consider the shared memory (thread based) parallelization of a dense MVP $y=Ax$. The shared data that the run-time system needs to keep track of in order to ensure a correct end result is the data that will be modified during the execution. In the example, this is only the output vector $y$. To implement the multiplication as a task parallel algorithm, we need to break down the operation into smaller components. This can be done in several different ways by blocking or slicing the matrix and vectors into smaller partitions. 

\begin{figure}[!htb]
\centering
\scalebox{.4}{
\boldmath%
\begin{tikzpicture}[line width=2pt]
\begin{scope}
\fill[white] (0,0) rectangle (.5,4);
\fill[gray] (0,2) rectangle (.5,3);
\draw (0,0) grid (.5,4);
\draw (0,0) rectangle (.5,4);
\end{scope}
\node at (1.7,1.95) [text width=2cm,font=\Huge] {$\leftarrow$};
\begin{scope}[xshift=55]
\fill[white] (0,0) rectangle (4,4);
\fill[gray] (1,2) rectangle (2,2+1);
\draw (0,0) grid (4,4);
\end{scope}
\begin{scope}[xshift=180]
\fill[gray] (0,2) rectangle (.5,3);
\draw (0,0) grid (.5,4);
\draw (0,0) rectangle (.5,4);
\end{scope}
\begin{scope}[xshift=-240]
\fill[white] (0,0) rectangle (.5,4);
\fill[gray] (0,0) rectangle (.5,4);
\draw (0,0) rectangle (.5,4);
\end{scope}
\begin{scope}[xshift=-240]
\node at (1.7,1.95) [text width=2cm,font=\Huge] {$\leftarrow$};
\end{scope}
\begin{scope}[xshift=55-240]
\fill[white] (0,0) rectangle (4,4);
\fill[gray] (1,0) rectangle (2,4  );
\draw (0,0) rectangle (1,4);
\draw (1,0) rectangle (2,4);
\draw (2,0) rectangle (3,4);
\draw (3,0) rectangle (4,4);
\end{scope}
\begin{scope}[xshift=180-240]
\fill[gray] (0,2) rectangle (.5,3);
\draw (0,0) grid (.5,4);
\draw (0,0) rectangle (.5,4);
\end{scope}
\begin{scope}[xshift=240]
\fill[white] (0,0) rectangle (.5,4);
\fill[gray] (0,2) rectangle (.5,2+1);
\draw (0,0) grid (.5,4);
\draw (0,0) rectangle (.5,4);
\end{scope}
\begin{scope}[xshift=240]
\node at (1.7,1.95) [text width=2cm,font=\Huge] {$\leftarrow$};
\end{scope}
\begin{scope}[xshift=55+240]
\fill[white] (0,0) rectangle (4,4);
\fill[gray] (0,2) rectangle (4,2+1);
\draw (0,0) rectangle (4,1);
\draw (0,1) rectangle (4,2);
\draw (0,2) rectangle (4,3);
\draw (0,3) rectangle (4,4);
\end{scope}
\begin{scope}[xshift=180+240]
\fill[gray] (0,0) rectangle (.5,4);
\draw (0,0) rectangle (.5,4);
\end{scope}
\end{tikzpicture}}
\caption{Three different partitionings of the matrix and vectors, and the responsibility of a single task of the MVP. The data accessed by the task is shaded.}
\label{fig:blocking}
\end{figure}
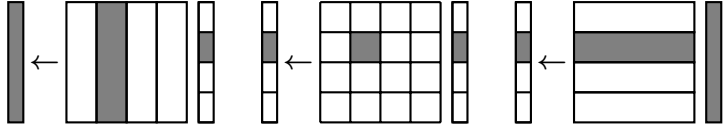

Figure~\ref{fig:blocking} shows three common ways of splitting the data structures. A task would correspond to multiplying one slice or block of the matrix with the corresponding part of the vector $x$. The column-based slicing of the matrix is a bad choice because the whole output vector is touched by each task, and the tasks can therefore not run in parallel (without modifications). The block partitioning as well as the row slicing scheme both allow for parallelism as only a part of the shared data $y$ is touched. The latter two provide the same level of parallelism, but if the multiplication is part of a larger scheme where there is a possibility to interleave tasks from different operations, the block partitioning may be preferred because it provides a larger number of tasks.

The number of tasks, especially in relation to the size of the tasks, is important for performance. With too few tasks, the amount of parallelism is reduced, and there is not enough work for all the threads, which leads to idle time. If there is instead a large number of tasks, but these are very small in terms of computational work, the overhead from managing the tasks in the run-time system becomes large in relation to the task size. Optimally, the task sizes should be chosen such that the task data fits within the local cache. 

A slightly different problem that also has a significant impact on performance is bandwidth contention. That is, when all threads are trying to fetch data for their tasks, there may not be enough bandwidth to supply the threads at full speed. Hence, the parallel execution will not be as efficient as theoretically expected. There are different ways to counter bandwidth related problems. If possible, similar computations should be combined such that the number of floating point operations per memory access is increased. This was successfully used, e.g., in ~\cite{TiLaLeFly15}. If there is a mix of tasks, with and without bandwidth sensitivity, resource-aware scheduling can improve performance~\cite{TiLaBaMa15}. Finally, accessing contiguous memory locations is more efficient than random memory accesses, which means that data structures should be allocated in such a way that tasks read from contiguous memory as far as possible. 











\section{Parallel implementation of the fast MVP}\label{sec:impl}
In a typical application, the MVP is performed a large number of times within an iterative method while the matrix structure is built once and can be seen as static. Therefore, we focus on the multiplication algorithm even if the build step could also be parallelized. In the following subsections, we will discuss the algorithm at a general level and the steps taken to convert it to a task parallel implementation.

\subsection{The properties of the algorithm}






Technically, the hierarchical NESA MVP algorithm consists of a large number of smaller dense MVPs. What is interesting from the parallelization perspective is the number of operations, the size of the operations, and the dependency structure. In order to discuss concrete numbers, we denote the number of active group at each level by $N_\ell$. For a full tree in $d$ dimensions $N_{\ell+1}=2^dN_\ell$, but here the tree is sparse. For the example we are using, the average number of children is around 2.5 leading to $N_{\ell+1}\approx 2.5N_\ell$. We choose to express all numbers of operation in terms of $N_L$. Then we have $N_\ell=N_L/2.5^{(L-\ell)}$.


Going back to the algorithm described in Section~\ref{sec:model} by equations~\eqref{eq:algnf}--\eqref{eq:algrec}, we make the following observations:

\begin{itemize}
\item The matrices in the near field operation~\eqref{eq:algnf} are on average of size $P\times P$ but the sizes vary between the groups. Each group is involved in at most $3^d$ near field operations, but in our sparse tree, the average is 5 including the self interaction. The total number of near field tasks is then approximately $5N_L$. These operations can be performed in any order, but updates to the same vector $[s]_\beta$ must be performed one at a time.

\item The radiation~\eqref{eq:algrad} and reception operations~\eqref{eq:algrec} are completely independent. The matrix sizes are on average $Q\times P$ and $P\times Q$, respectively, and the number of operations is $N_L$ for each type.

\item The number of source transfer~\eqref{eq:algsou} and potential transfer operations~\eqref{eq:algpot} are given by the total number of children as each parent and child interact once. We have $N_c=\sum_{\ell=\ell_0+1}^{L}N_\ell\approx N_L\sum_{j=0}^{L-(\ell_0+1)}2.5^j$ leading to $
N_c\leq\frac{5}{3}N_L$.
Due to the equivalent source formulation, all the transfer matrices are of the same size, $Q\times Q$. These operations are ordered in the sense that children and parents must complete their tasks in the correct order. Also, when the children are contributing to their parent's equivalent sources, updates by different children must be performed one at a time. Otherwise, operations at the same level are independent. 

\item The translation operations~\eqref{eq:algtrans} also involve matrices of size $Q\times Q$. The number of groups in the interaction lists at each level is limited. In our application, the average number is less than 7 for all levels.
This means that the number of translation tasks is less than $7\sum_{\ell=\ell_0}^LN_\ell\leq 35/3N_L$. The translation operations can be performed in any order, but updates to the same location must be performed one at a time.
\end{itemize}

The hierarchical matrix representation based on groups and levels provides a natural description of the algorithm in terms of tasks. The number of tasks is large, and even if there are dependencies between the levels in the algorithm, there are plenty of tasks that are independent, and can be interleaved with the constrained tasks. In particular, it should be of benefit to mix the independent near-field and dependent far-field computations.

The sizes of the constituent matrix multiplications are completely determined by the parameters $P$ and $Q$. However, if we modify $P$ or $Q$ for a fixed problem, we change the total amount of work in the algorithm as well as the memory requirements. By changing $Q$, we also change the accuracy of NESA. This means that we cannot easily optimize the task sizes. It also means that the number of source points $N$ will not have a strong influence on the performance. For very small $N$, the number of tasks may be too small to provide work for all threads in parallel, but as soon as $N$ is large enough, the performance will be determined by the task sizes.

For our two-dimensional test problem, the preferred values of $P$ and $Q$ are small with, e.g., $Q=10$ leading to a tolerance of around $1e-5$. In the three-dimensional case, when the auxiliary sources are placed on a sphere instead of a circle, the corresponding numbers are larger, for a similar accuracy $Q\approx 100$ is needed~\cite{2014_Li_ETAL_Nested}. When the tasks are too small, the overhead from managing the tasks dominate and the potential parallel speedup is reduced or lost. The precise sizes needed to achieve speedup will be investigated in section~\ref{sec:perf}.

One way to avoid the overhead resulting from having too small tasks would be to let each task manage several groups. This would reduce the number of tasks, but it would increase the complexity of the algorithm, making it harder to implement.

Another performance issue that the MVP will suffer from is bandwidth contention. All MVPs are somewhat sensitive to bandwidth as the number of floating point operations $\mathcal{O}(N^2)$ are proportional to the number of matrix elements $\mathcal{O}(N^2)$. If the operations are instead transformed into matrix--matrix operations, the situation is improved, as the number of operations become $\mathcal{O}(N^3)$, while the storage is still $\mathcal{O}(N^2)$. In the NESA algorithm this would be possible for the radiation and reception steps, since the transfer matrices between parents and children with the same relative positions are the same. Then all similar products could be combined into one operation. However, this would also make the implementation much more complicated with additional packing and unpacking procedures, which also could be time consuming.










\subsection{The task parallel implementation}
We have chosen to implement the most natural task-based formulation, where each task corresponds to one small MVP. In the sequential code, each of the small matrices are allocated in contiguous memory using a C++ user defined \texttt{Matrix} data type, and the MVPs are performed by directly calling the BLAS routine \texttt{cblas\_dgemv}. The \texttt{Matrix} type is used also for the input and output vectors.

The changes that are needed to produce a task parallel code are minor. First we need to protect the output vectors from simultaneous accesses by different tasks. In SuperGlue, shared data is protected by handles that control accesses to the data. Therefore, we introduce a new type \texttt{SGMatrix}, which basically equips a \texttt{Matrix} with a handle.
The type definition is shown in~\ref{app:A}.

Next we need to, unless it is already available, define a SuperGlue task class that provides an MVP. The class is also shown in~\ref{app:A}. When the task class is in place, we construct a new \texttt{gemv} subroutine that instead of directly calling BLAS, constructs the corresponding task and submits it to the run-time system. The task parallel \texttt{gemv} subroutine is shown as Program~\ref{prg:gemv}.
\begin{lstlisting}[numbers=left,numbersep=2pt, numberstyle=\tiny,
numberfirstline=true,language=C++,
caption={The subroutine that submits an MVP task.},
label=prg:gemv,mathescape]
void gemv(SGMatrix &A, SGMatrix &x,SGMatrix &y){
    SGTaskGemv *t= new SGTaskGemv(A,x,y);
    sgEngine->submit(t); 
}
\end{lstlisting}

We can now write the whole program in task parallel form, by replacing the data types and the subroutine calls by their counterparts. Program~\ref{prg:main} shows how SuperGlue is invoked, how matrices are created and how one MVP is performed.
\begin{lstlisting}[numbers=left,numbersep=2pt, numberstyle=\tiny,
numberfirstline=true,language=C++,
caption={An example of how SuperGlue is included in a program performing an MVP.},
label=prg:main,mathescape]
#include "superglue.hpp"

SuperGlue<Options> *sgEngine;

int main(int argc , char *argv[]){
  // Start the task parallel run-time system
  sgEngine = new SuperGlue<Options>(config.cores);
  // Allocating matrices (filled with 0.0)
  int P=300, Q=100;
  Matrix &a = * new Matrix (Q,P,0.0);
  Matrix &x = * new Matrix (P,1,0.0);
  Matrix &y = * new Matrix (Q,1,0.0);
  // Make these protected shared data
  SGMatrix &A = *new SGMatrix(a);
  SGMatrix &X = *new SGMatrix(x);
  SGMatrix &Y = *new SGMatrix(y);
  // Write the algorithm with calls to the MVP 
  // subroutine. For each new call, the 
  // corresponding task is submitted
  gemv(A,X,Y);
  // Wait for all tasks to finish
  sgEngine->barrier();
}
\end{lstlisting}

As mentioned above, more details on the task class implementation and the shared data type is given in~\ref{app:A}.


\subsection{OpenMP implementation}\label{sec:ompimp}
An implementation, with a similar functionality as the task parallel implementation described above, can with some care be created also with OpenMP. A simple \textit{task} construct was introduced in OpenMP 3.0, and a \textit{depend} clause was added in OpenMP 4.0, to allow dependencies between sibling tasks, i.e, tasks created within the same parallel region. This means that if we create several parallel regions for different parts of the algorithm, there will effectively be barriers in between, and the tasks from different regions cannot mix. 

The proper way to do it is to create one parallel region that covers the whole computation, and then make sure that only one thread generates tasks such that the sequential order is not compromised. An excerpt from the OpenMP main program that illustrates this is shown in Program~\ref{prg:ompmain}. The tasks are submitted from the two subroutines that are called.
\begin{lstlisting}[numbers=left,numbersep=2pt, numberstyle=\tiny,
numberfirstline=true,language=C++,
caption={The global parallel region in the main program using OpenMP.},
label=prg:ompmain,mathescape]
#pragma omp parallel 
    {
      #pragma omp single
      {
	// Submit tasks for near-field multiplication
	FMM::mv_near_field(OT,C,Q);                 
	// Submit tasks for far-field multiplication
	FMM::mv_far_field(OT,C,Q);
      }
    }
#pragma omp taskwait 
#pragma omp barrier 
\end{lstlisting}

The tasks are defined using the task pragma with the depend clause. Only the (necessary) inout dependence for the output data vector is included. Adding the (nonessential) read dependencies on the matrix and input data vector was shown in the experiments to degrade performance.  
\begin{lstlisting}[numbers=left,numbersep=2pt, numberstyle=\tiny,
numberfirstline=true,language=C++,
caption={The OpenMP task pragma that defines a gemv task.},
label=prg:omptask,mathescape]
#pragma omp task depend(inout:Y[0:N])
cblas_dgemv(COL_MAJOR,transA,M, N, 1.0, Mat, lda, X, 1, 1.0, Y, 1);
\end{lstlisting}

As can be seen, the implementation is not so difficult, but there are several ways to make mistakes that lead to suboptimal performance. The programmer needs to understand how the task generation, the task scheduling, and the parallel regions interact.
\section{Performance evaluation}\label{sec:perf}
The experiments have been performed on one shared memory node of the Tintin 
cluster at the Uppsala Multidisciplinary Center for Advanced Computational Science (UPPMAX). Each node is dual socket with two AMD Opteron 6220 (Bulldozer) processors running at 3.0 GHz with 64 GB or 128 GB memory. A peculiarity of the Bulldozer architecture is that each floating point unit (FPU) is shared between two cores. This means that the theoretical speedup when using $2p$ threads (cores) is only $p$, and the highest theoretical speedup on one node with 16 threads is 8.

The same problem with $N=100\,000$ source points is solved in all experiments, but the method parameters $P$ (the average number of source points at the finest level) and $Q$ (the number of auxiliary points used for each group) are varied between the experiments.

As discussed in the previous section, the properties of the near and far field parts of the algorithm are quite different. Here, we first evaluate the two parts separately to establish their performance at different parameter choices, and then move to the full MVP algorithm.

For each test case, we show speedup results computed as
\begin{equation}
S_p = \frac{T_1}{T_p},
\end{equation}
where $T_p$ is the execution time of the task parallel implementation running on $p$ threads. In the graphs, we also show the speedup of the sequential code compared with the parallel code running on one thread. In the optimal case they would be close to equal, but for problems with small tasks, we can see a slight difference.

We also show execution traces, where each task is shown as a triangle with its base at the starting time of the task and the tip at the end of that task's execution. Traces provide a good way of visualizing the execution, the scheduling, and potential performance issues.

Finally, we provide tables with detailed information on execution times, speedup and utilization. We define the utilization as
\begin{equation}
U_p = \frac{T_p^t}{T_p},
\end{equation}
where $T_p^t$ is the fraction of the execution time spent executing tasks. The remaining time represents the overhead of managing tasks, idle time when threads are waiting for tasks, and load imbalance at the end of the execution.









\subsection{The far field computation}
\begin{figure}[!htb]
\centering
\includegraphics[width=0.5\textwidth]{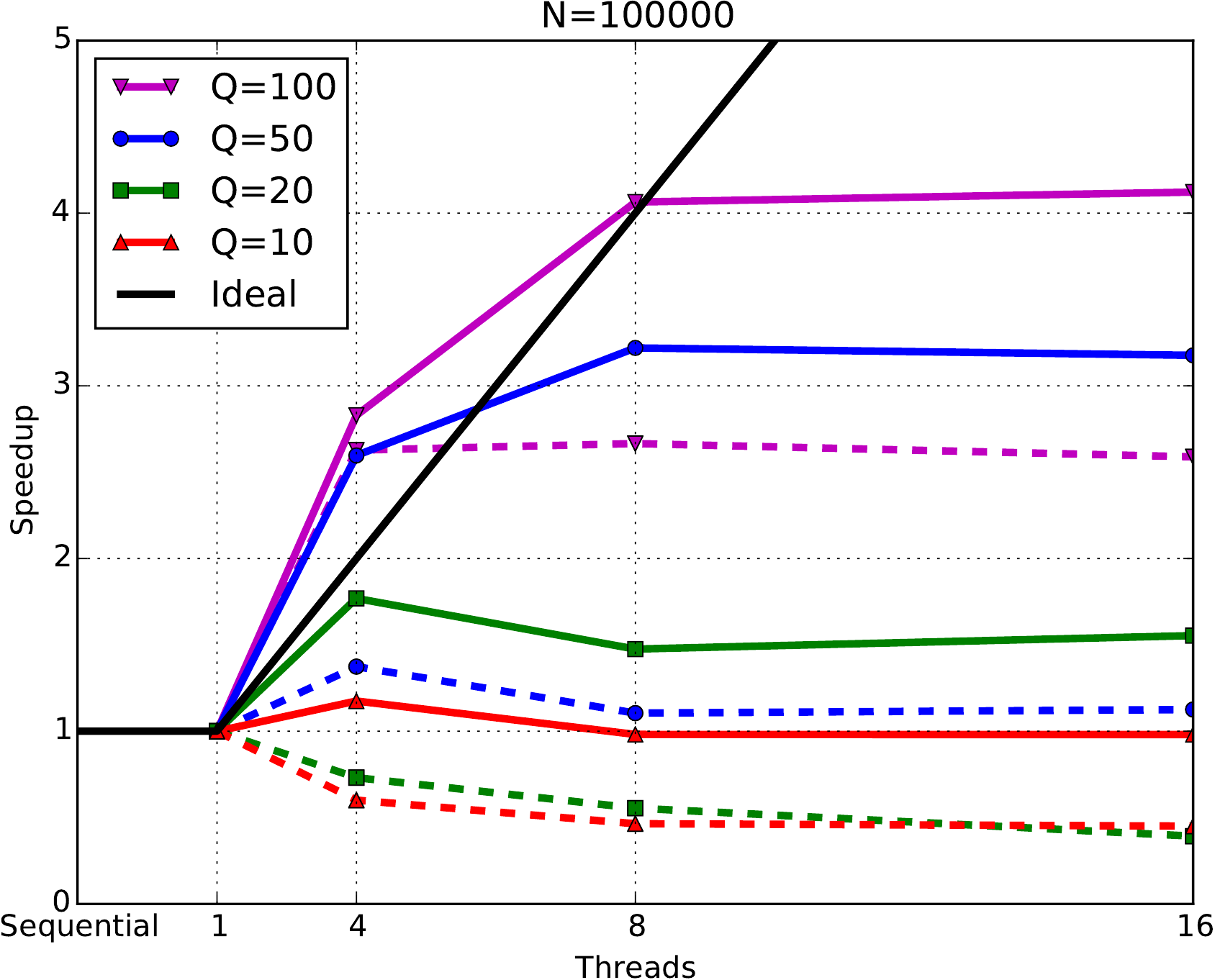}
\caption{Speedup for different values of $Q$ for the far field computation with $P=400$ (solid lines) and with $P=50$ (dashed lines).}
\label{fig:ffspeedup}
\end{figure}

Figure~\ref{fig:ffspeedup} shows how the speedup of the far field computations varies with $Q$ for two different values of $P$. The speedup grows with $Q$ as the task sizes increase, and it also grows with $P$, which affects the size of the radiation and reception tasks. However increasing only $Q$ is not enough, which tells us that transfer and translation tasks do not scale very well.



\begin{figure}[!htb]
\centering
\includegraphics[width=\textwidth,height=0.3\textwidth]{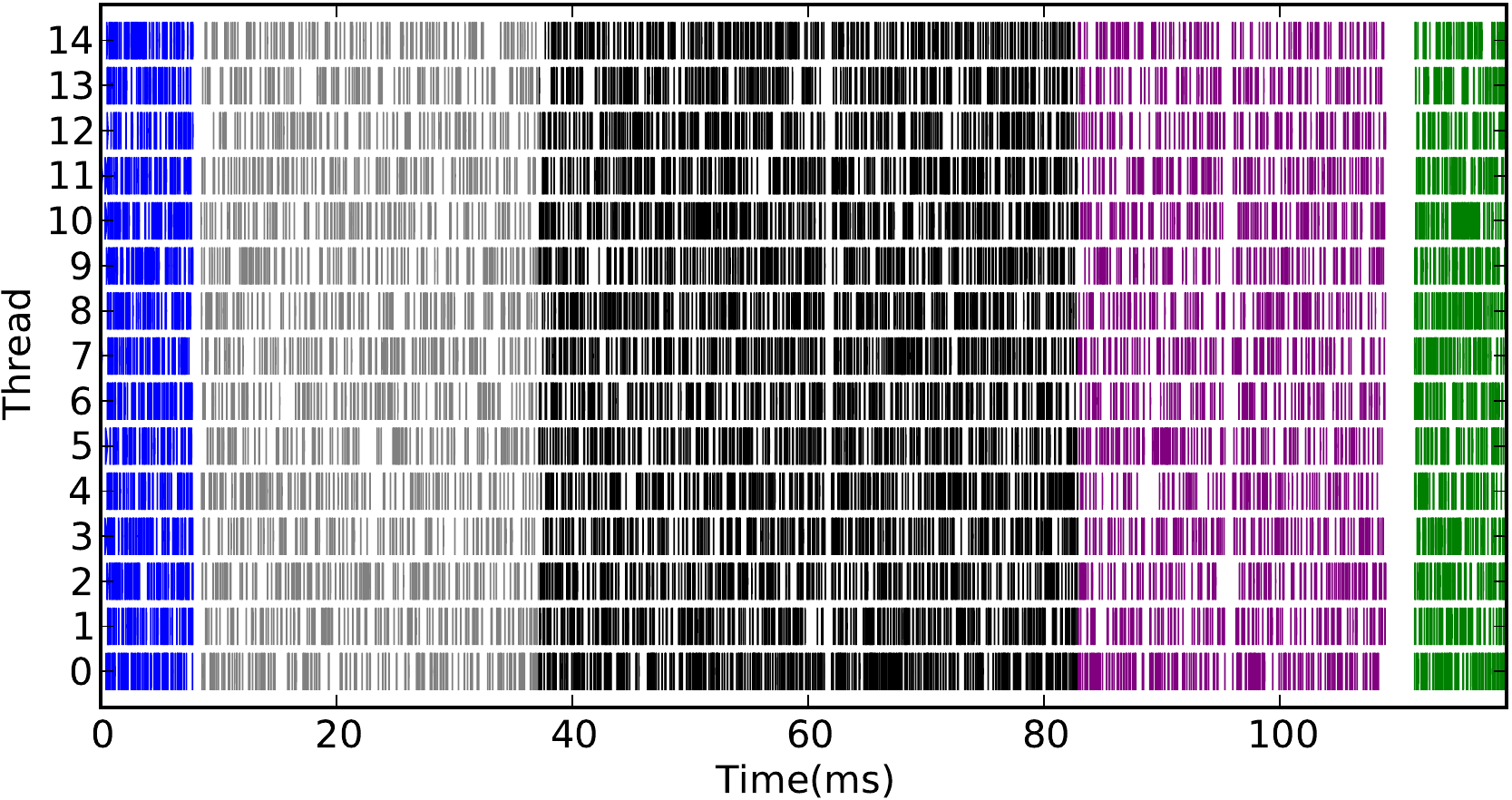}
\includegraphics[width=\textwidth,height=0.3\textwidth]{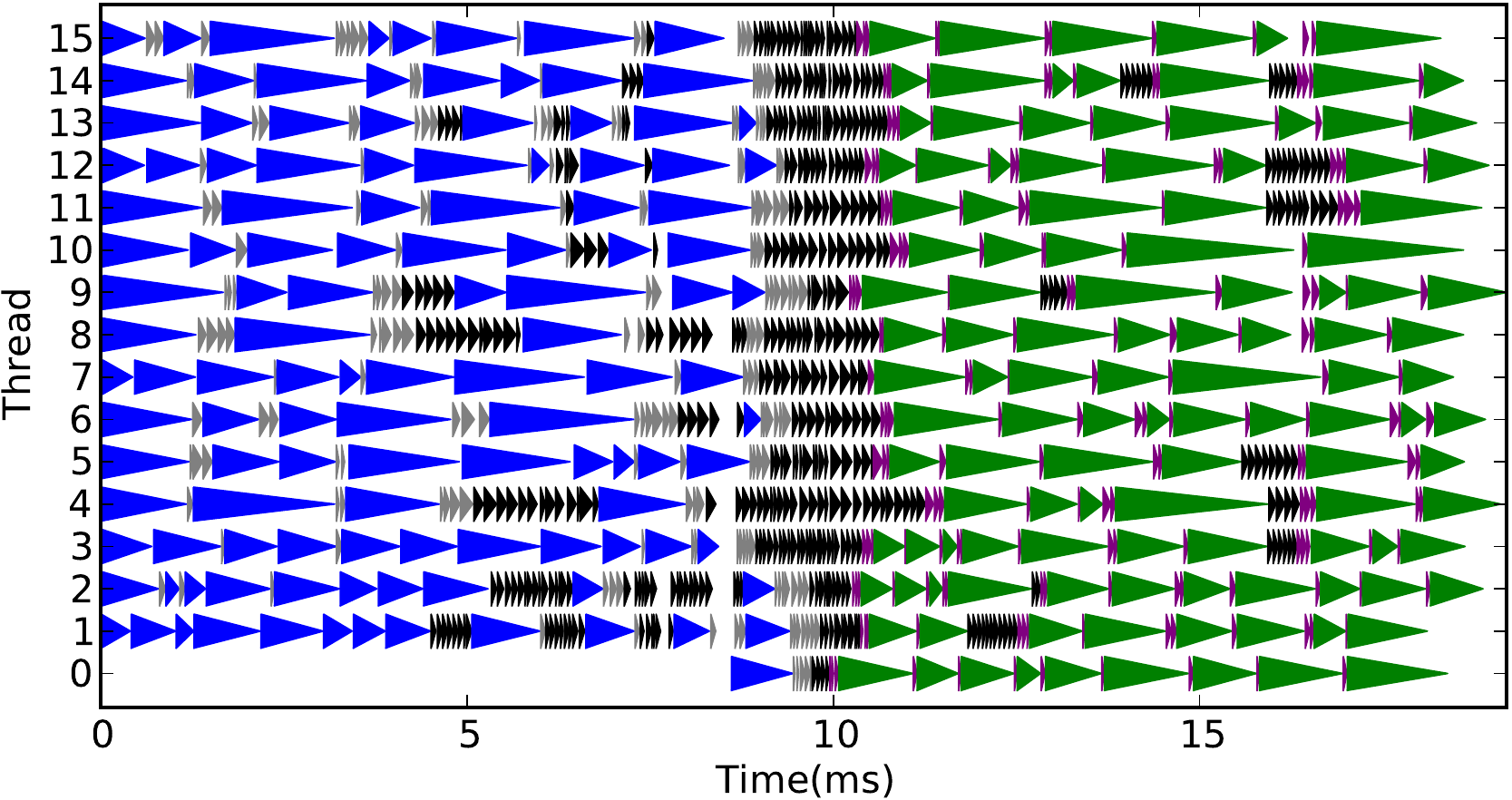}
\caption{Execution traces for the far field computation for $Q=10$ and $P=50$ (top) and for $Q=100$ and $P=400$ (bottom).}
\label{fig:traceQ}
\end{figure}

To learn more about the details of the execution, we look at the traces for execution on 16 threads shown in Figure~\ref{fig:traceQ}. For the problem with small task sizes there are several things to observe. First, the colors that represent different types of tasks are not mixed. This tells us that the tasks are so small that they finish executing before more tasks have been submitted. In fact, only 15 threads are visible in the trace because one thread is constantly occupied with task submission. The small size of the tasks also leads to idle time in between tasks and the resulting execution is not efficient.

For the problem with larger $P$ and $Q$, we see the benefit of having the larger radiation and reception tasks. These are completely independent and provide enough work to occupy all threads, and the troublesome transfer tasks are nicely embedded within these computations. The translation tasks are smaller, but also more independent than the transfer tasks, and they are mostly scheduled densely within the trace. The task submission is visible also here, but it only occupies the first half of the execution time for thread 0.

\begin{figure}[!htb]
\centering
\includegraphics[width=0.49\textwidth]{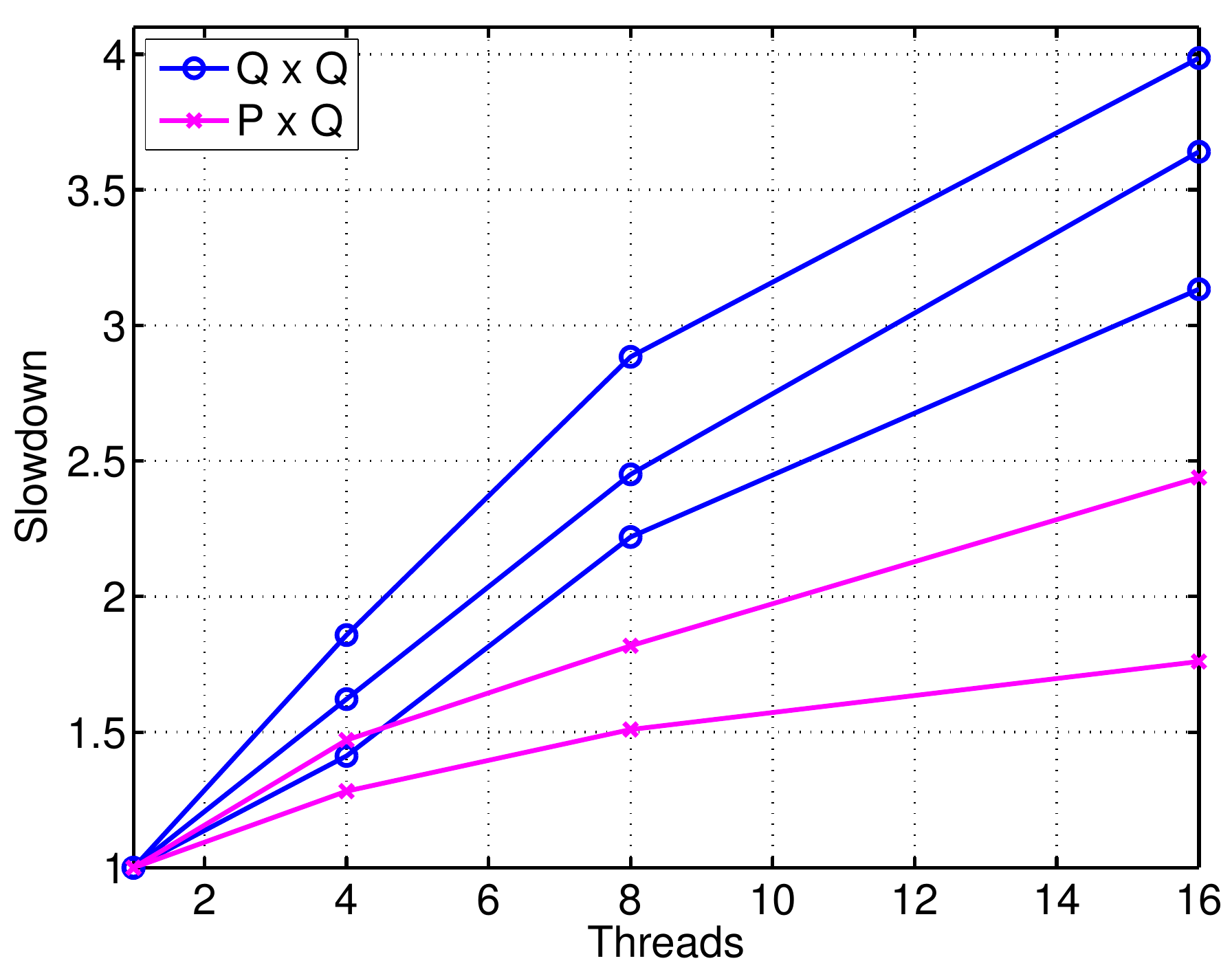}
\includegraphics[width=0.49\textwidth]{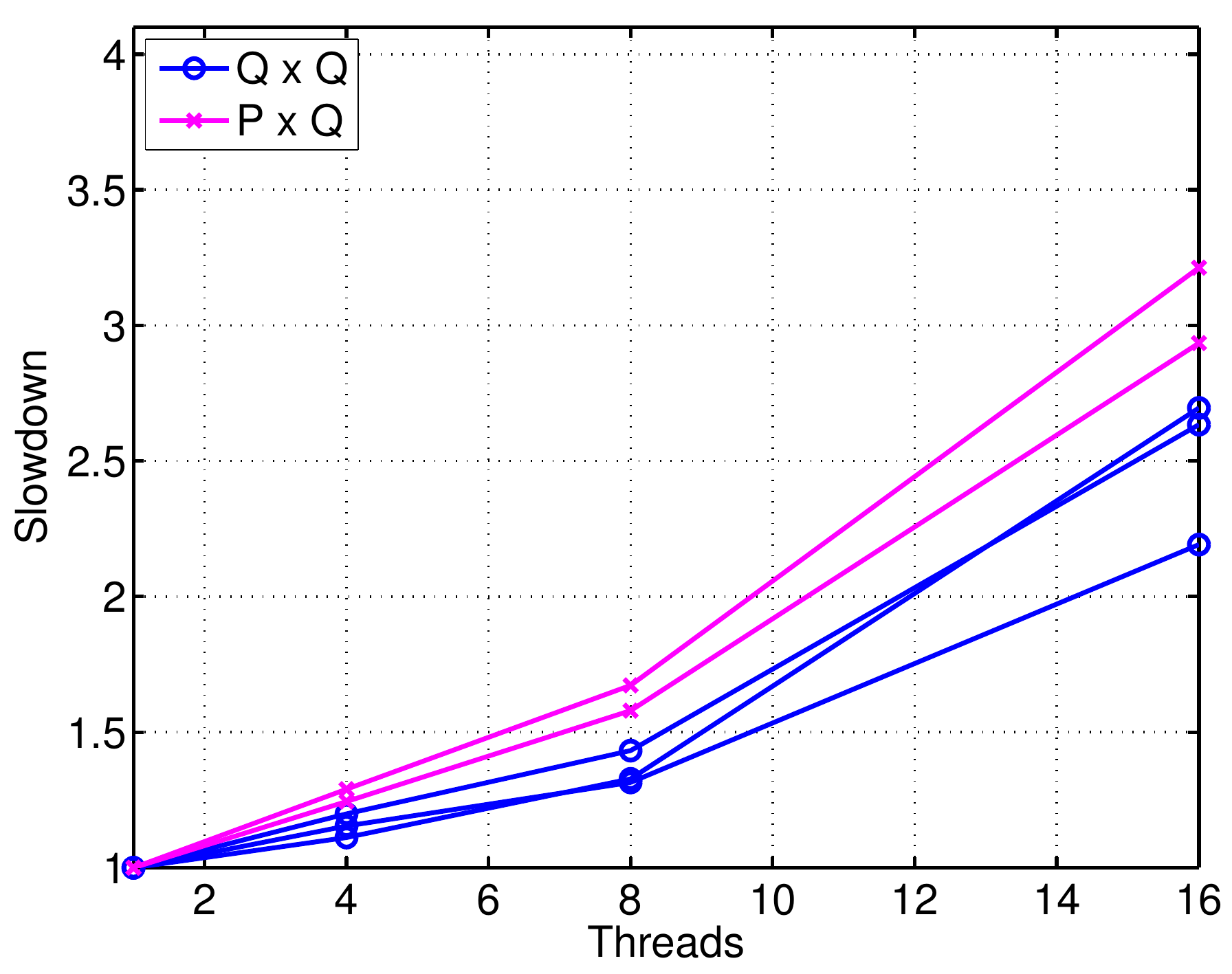}
\caption{Increase in individual task execution times due to resource contention for $P=50$, $Q=10$ (left) and $P=400$, $Q=100$ (right).}
\label{fig:slowdown}
\end{figure}

Looking at the second trace, we might expect a near optimal speedup as the schedule has very little idle time. However, the speedup was never above 4 in any of the experiments. To investigate if bandwidth contention is involved, we have computed the average execution time for each type of task for different numbers of threads. The result is shown in Figure~\ref{fig:slowdown}. In the left subfigure, we can see that all of the small tasks experience a slowdown or longer execution time per task on more threads. The larger tasks fare better, and it also seems that tasks with transposed matrices perform better. With an increase of 4 in the individual task execution times as in the worst case here, it is not possible to get an overall speedup higher than 4 on 16 threads. We can also here find the explanation to why the speedup is larger than the theoretical best for 4 threads. When only one thread is used it has dedicated use of the FPU. When instead each FPU is occupied by two threads, the computations become relatively slower, and the bandwidth issue is somewhat improved. In the right subfigure, where the tasks are larger, the slowdown is reduced, but there is still contention.

Finally, Table~\ref{tab:ff} shows speedup and utilization for the far field computations. For the problem with small task sizes, both speedup and utilization are very low, while for the problem with larger sizes, the utilization is very good, but due to the slowdown of individual tasks, the maximum speedup stays around 4.
%
\begin{table}
\caption{Performance results for the far field computation for two different parameter sets. $T_p$ is the execution time for $p$ threads, $S_p$ is the speedup, $S_p^*$ is the theoretical optimal speedup, and $U_p$ is the utilization.}
\label{tab:ff}
\centering
\begin{tabular}{rrrrr}\hline
\multicolumn{5}{c}{$Q=10$, $P=50$}\\\hline
$p$ & $T_p$ [ms] & $S_p$ & $S_p/S_p^*$ & $U_p$\\\hline
1 &  27&  1  &  1.00 & 0.38\\
4 &  86& 0.31 &  0.16 & 0.04\\
8 & 113& 0.24&    0.06 &  0.02\\
16& 119& 0.23&    0.03 &  0.01\\\hline
\multicolumn{5}{c}{$Q=100$, $P=400$}\\\hline
1 & 95 & 1  & 1.00& 0.98 \\
4 & 31 & 3.0& 1.52& 0.92\\
8 & 20 & 4.8& 1.19& 0.91\\
16& 19 & 5.0& 0.62 & 0.90\\\hline
\end{tabular}
\end{table}

\subsection{The near field computation}
\begin{figure}[!htb]
\centering
\includegraphics[width=0.5\textwidth]{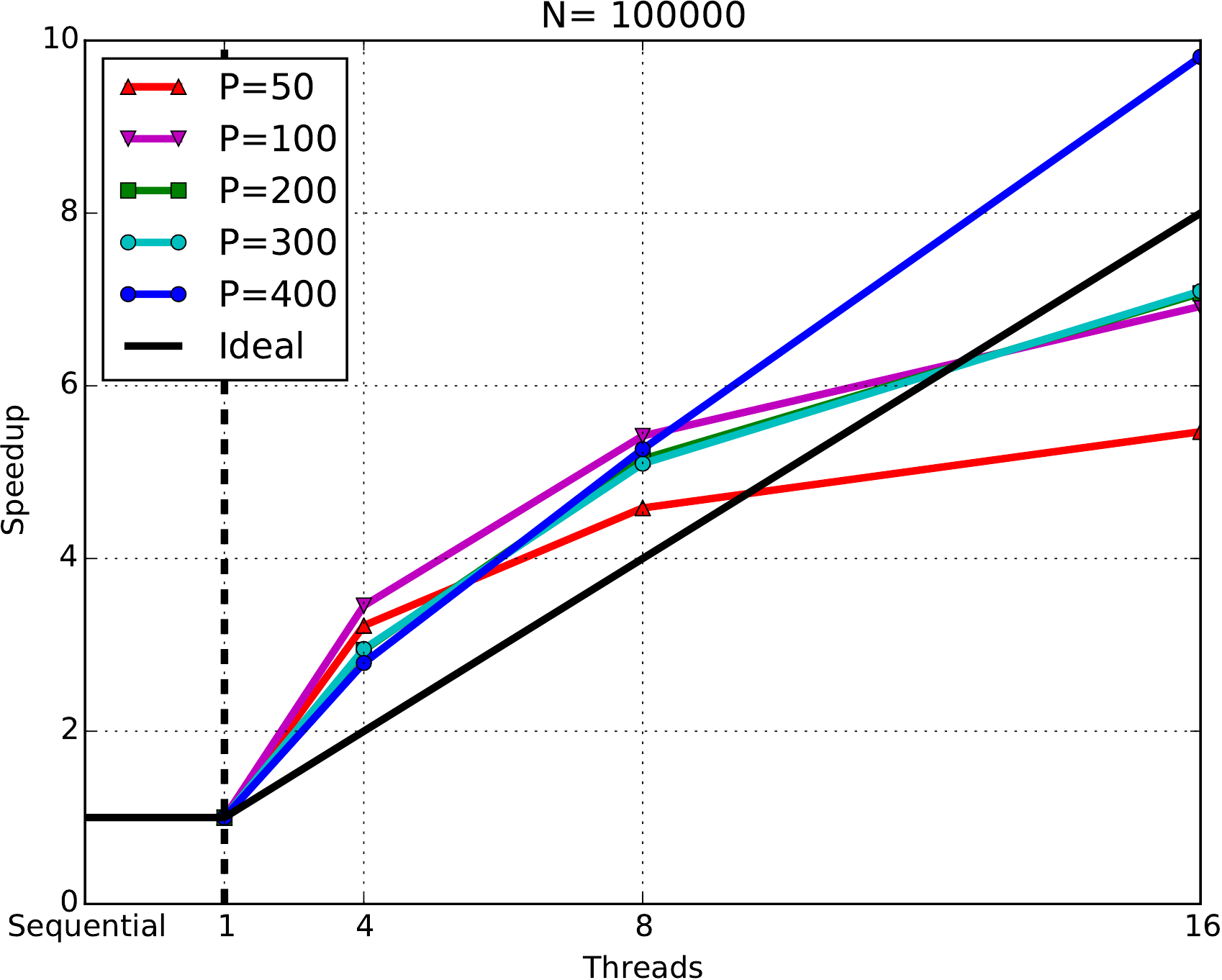}
\caption{Speedup for different values of $P$ for the near field computation.}
\label{fig:nfspeedup}
\end{figure}

For the near field computations, the picture is quite different. Speedup results are shown in Figure~\ref{fig:nfspeedup}. Also here, the speedup increases with $P$, but instead of leveling out at 4, it is superoptimal and lands at 10 for $P=400$.

\begin{figure}[!htb]
\centering
\hspace*{-3mm}\includegraphics[width=0.965\textwidth,height=0.3\textwidth]{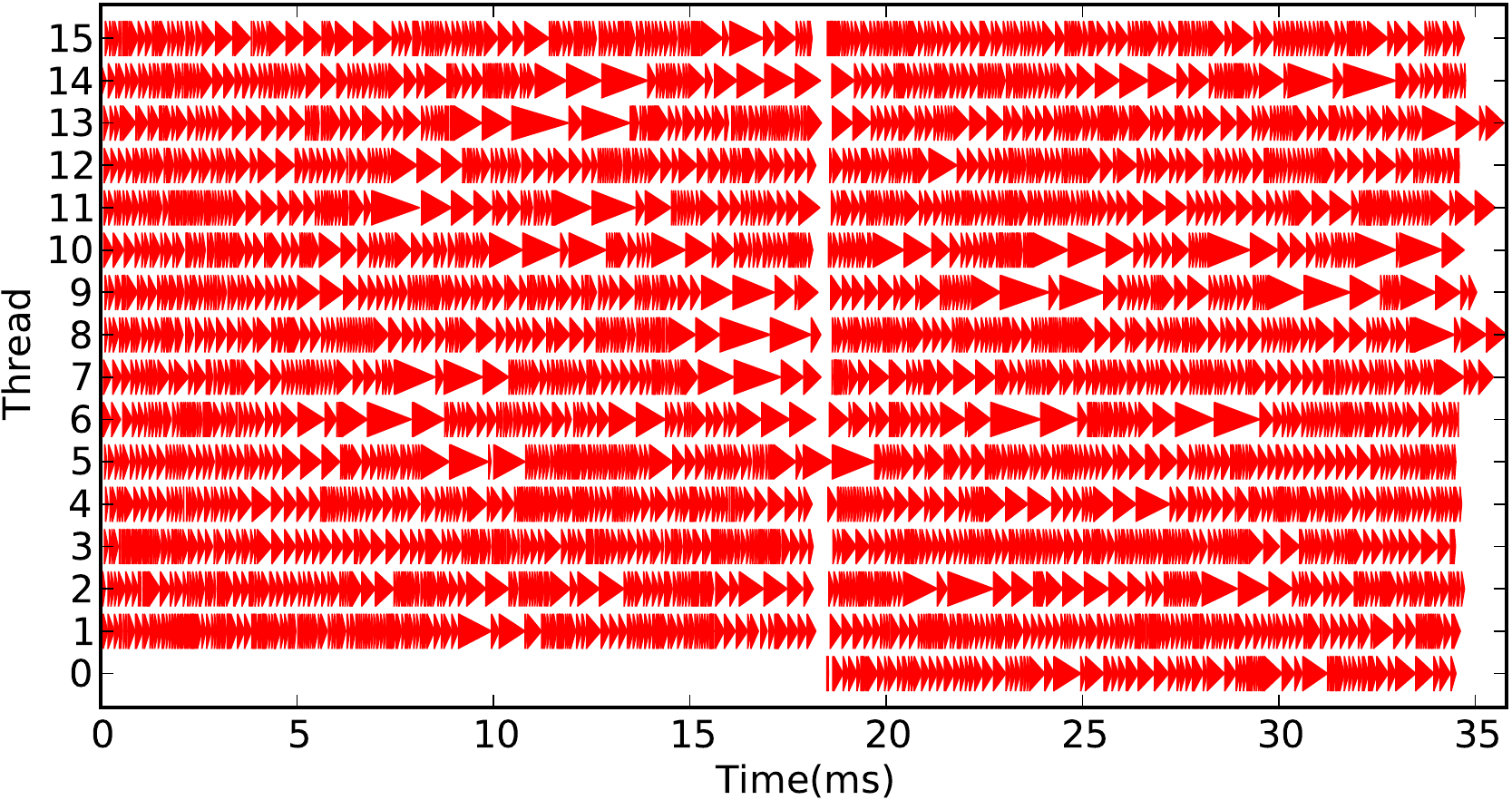}
\includegraphics[width=\textwidth,height=0.3\textwidth]{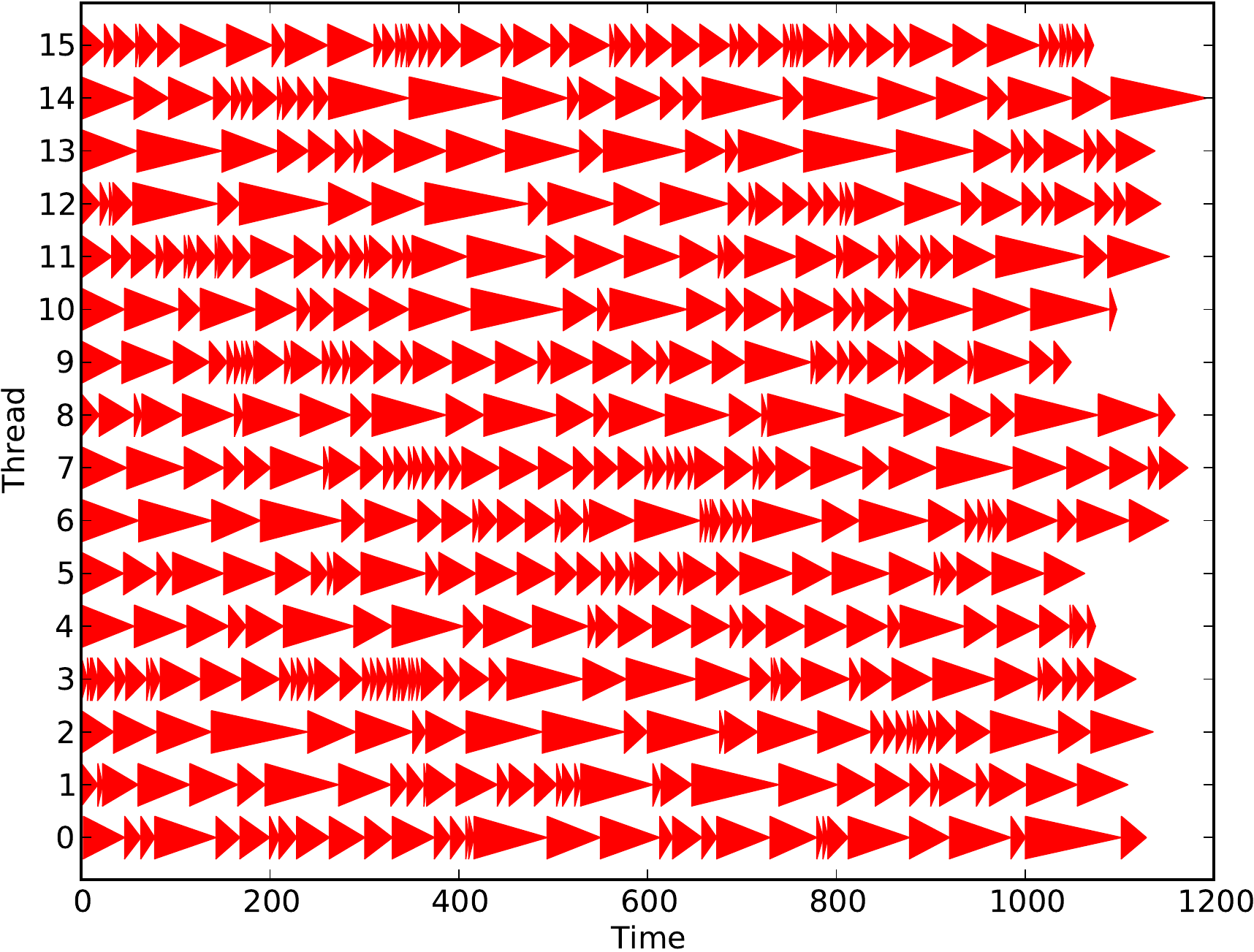}
\caption{Execution traces for the near field computation for $P=50$ (top) and for $P=400$ (bottom).}
\label{fig:traceP}
\end{figure}

The traces in Figure~\ref{fig:traceP} really show the strength of task parallel programming. The tasks are of highly varying sizes, depending on the number of source points in each group, but are scheduled densely across all threads. For the larger value of $P$, the task submission phase is not visible in the trace.

Table~\ref{tab:nf} shows performance results for the near field computation.
Already for the smaller $P$ both speedup and utilization are high, and for the larger $P$, the performance is excellent.
%
\begin{table}
\caption{Detailed performance results for the near field computation for two different parameter sets. $T_p$ is the execution time for $p$ threads, $S_p$ is the speedup, $S_p^*$ is the theoretical optimal speedup, and $U_p$ is the utilization.}
\label{tab:nf}
\centering
\begin{tabular}{rrrrr}\hline
\multicolumn{5}{c}{ $P=50$}\\\hline
$p$ & $T_p$ [ms] & $S_p$ & $S_p/S_p^*$ & $U_p$\\\hline
1 &  222&  1   &  1.00 & 0.97\\
4 &  66&   3.4 &  1.69 & 0.91\\
8 &  42&   5.2 &  1.30 & 0.89\\
16&  36&   6.2 &  0.77 & 0.87\\\hline
\multicolumn{5}{c}{$P=400$}\\\hline
1 & 3848 & 1  & 1.00& 1.00\\
4 & 1379 & 2.8& 1.40& 0.98\\
8 &  732 & 5.3& 1.31& 0.95\\
16&  398 & 9.7& 1.21 & 0.94\\\hline
\end{tabular}
\end{table}

\subsection{The complete MVP}
\begin{figure}[!htb]
\centering
\includegraphics[width=0.5\textwidth]{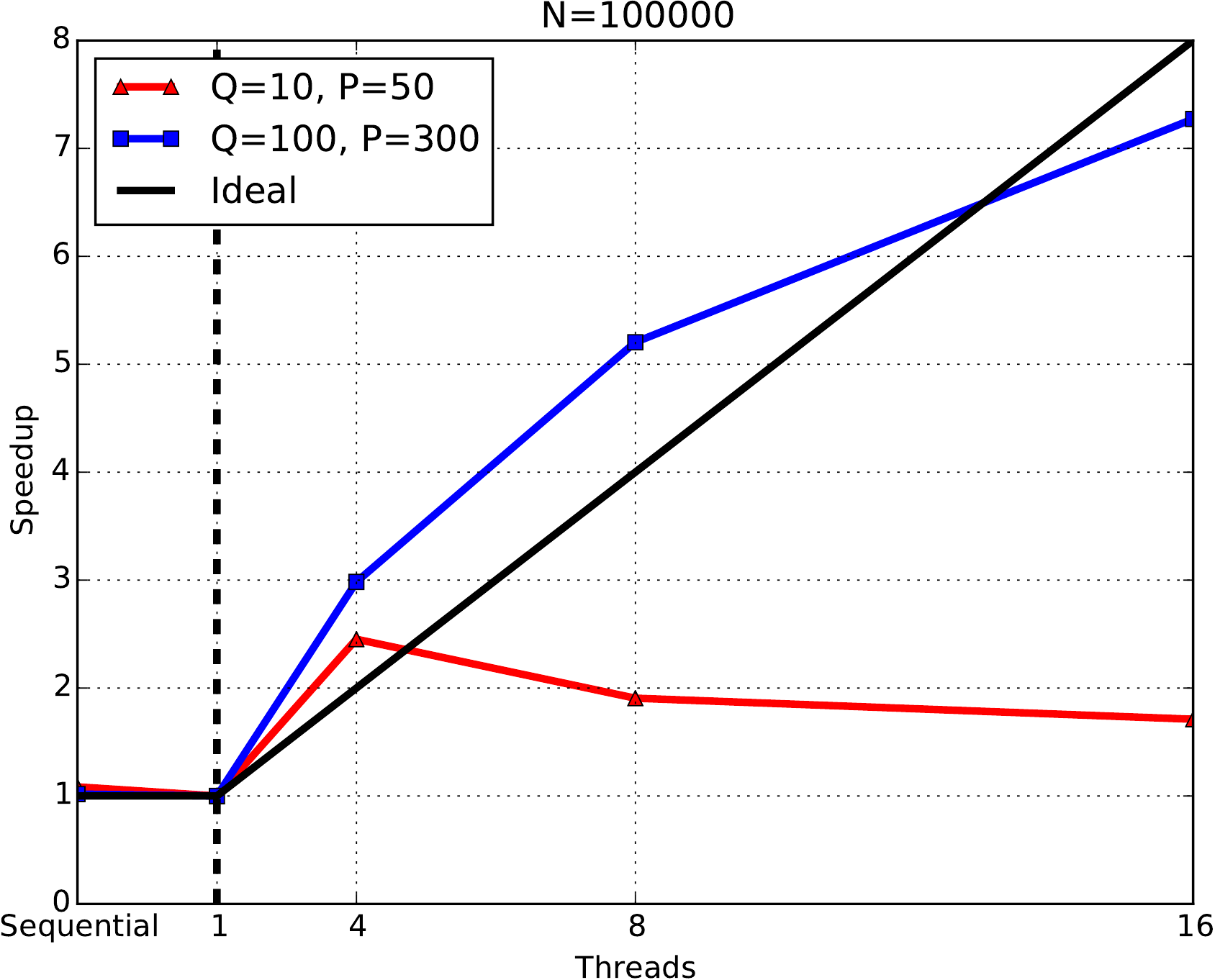}
\caption{Speedup for two combinations of $P$ and $Q$ resulting in different task sizes.}
\label{fig:speedup}
\end{figure}

Here we have run the complete MVP allowing the near field and far field tasks to mix when possible. Figure~\ref{fig:speedup} shows the speedup results for the complete runs. Here we have chosen to use $P=300$ for the problem with larger task sizes, even though the previous results showed that $P=400$ is more efficient. The reason is that we wanted the far field computation to be a larger part of the execution trace than for the $P=400$ case. For the problem with small tasks, the speedup is around 2, which is in between the far field result 0.23 and the near field result 6.2. For the problem with large tasks, the speedup 7.3 is very close to the result for the near field 7.45, and the far field result of 4.0 does not seem to impact the speedup at all.

\begin{figure}[!htb]
\centering
\includegraphics[width=\textwidth,height=0.3\textwidth]{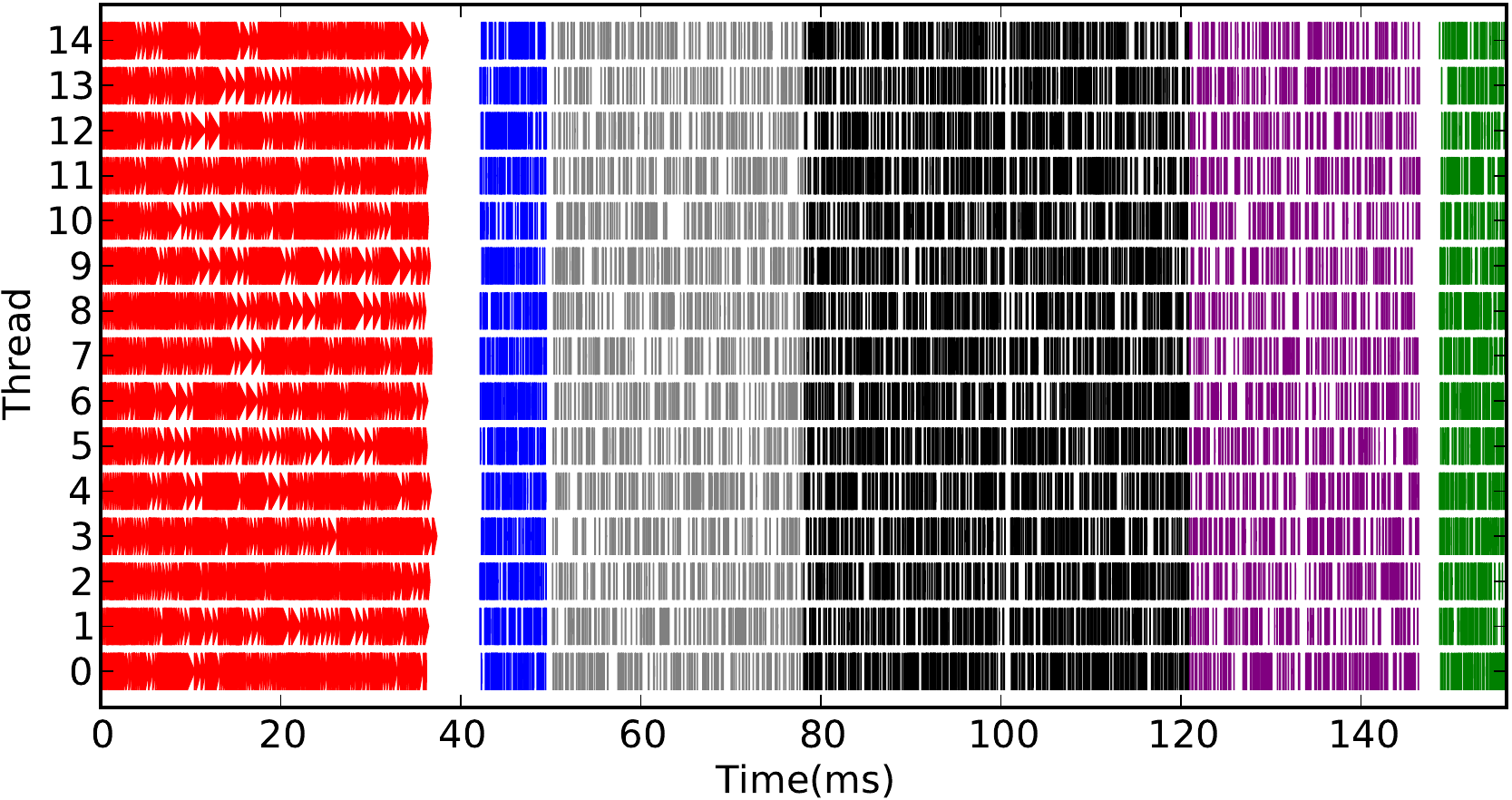}
\includegraphics[width=\textwidth,height=0.3\textwidth]{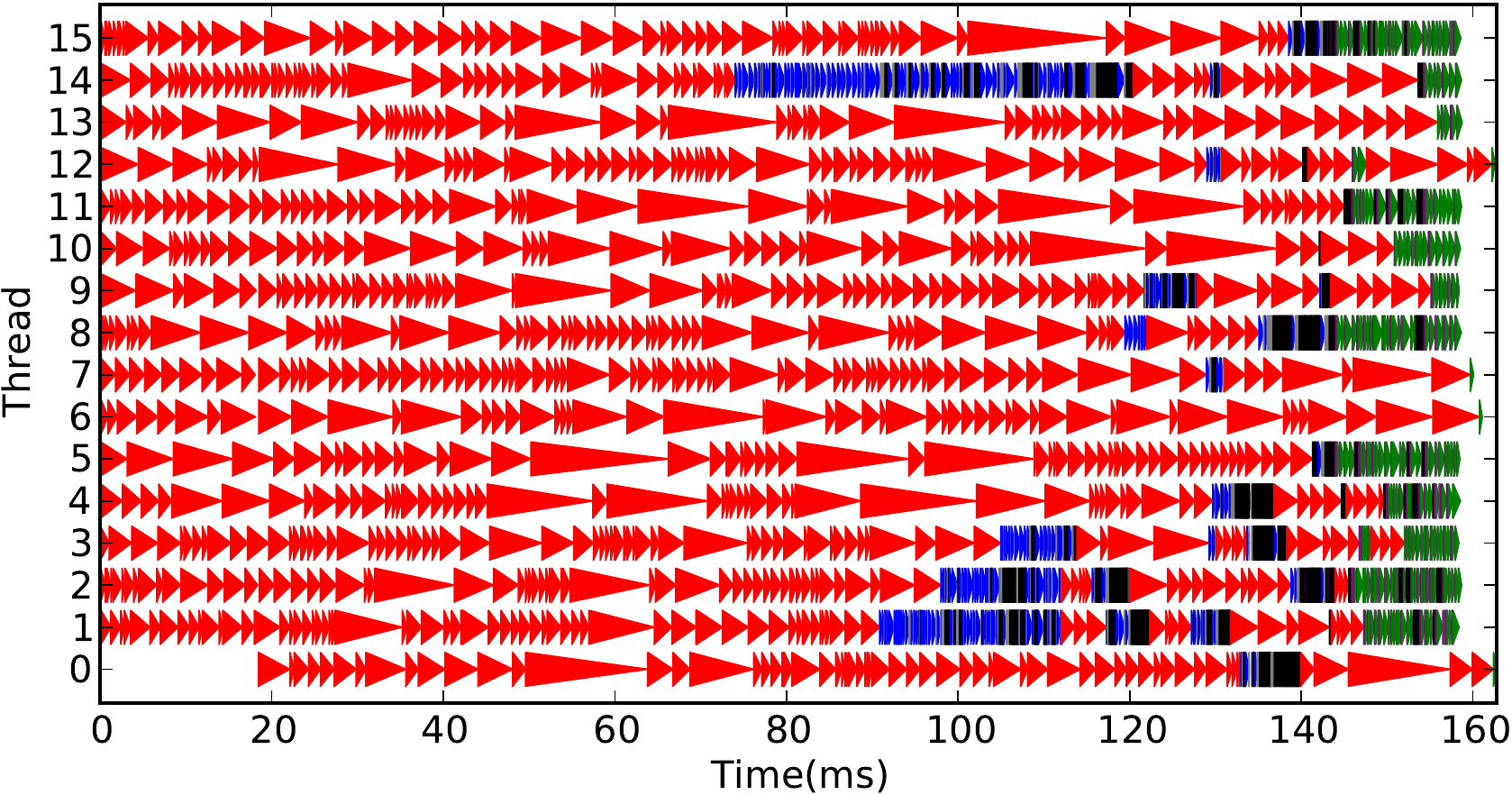}
\caption{Execution traces for the whole computation for $Q=10$ and $P=50$ (top) and for $Q=100$ and $P=300$ (bottom).}
\label{fig:tracePQ}
\end{figure}

To get further information, we consider the traces in Figure~\ref{fig:tracePQ}. In the case with small tasks, there is no mixing, and the speedup will clearly be a combination of that for the individual cases. However, when the tasks are larger, the executions are mixed and the far field computation is efficiently executed within the near field computation.

\begin{figure}[!htb]
\centering
\includegraphics[width=0.49\textwidth]{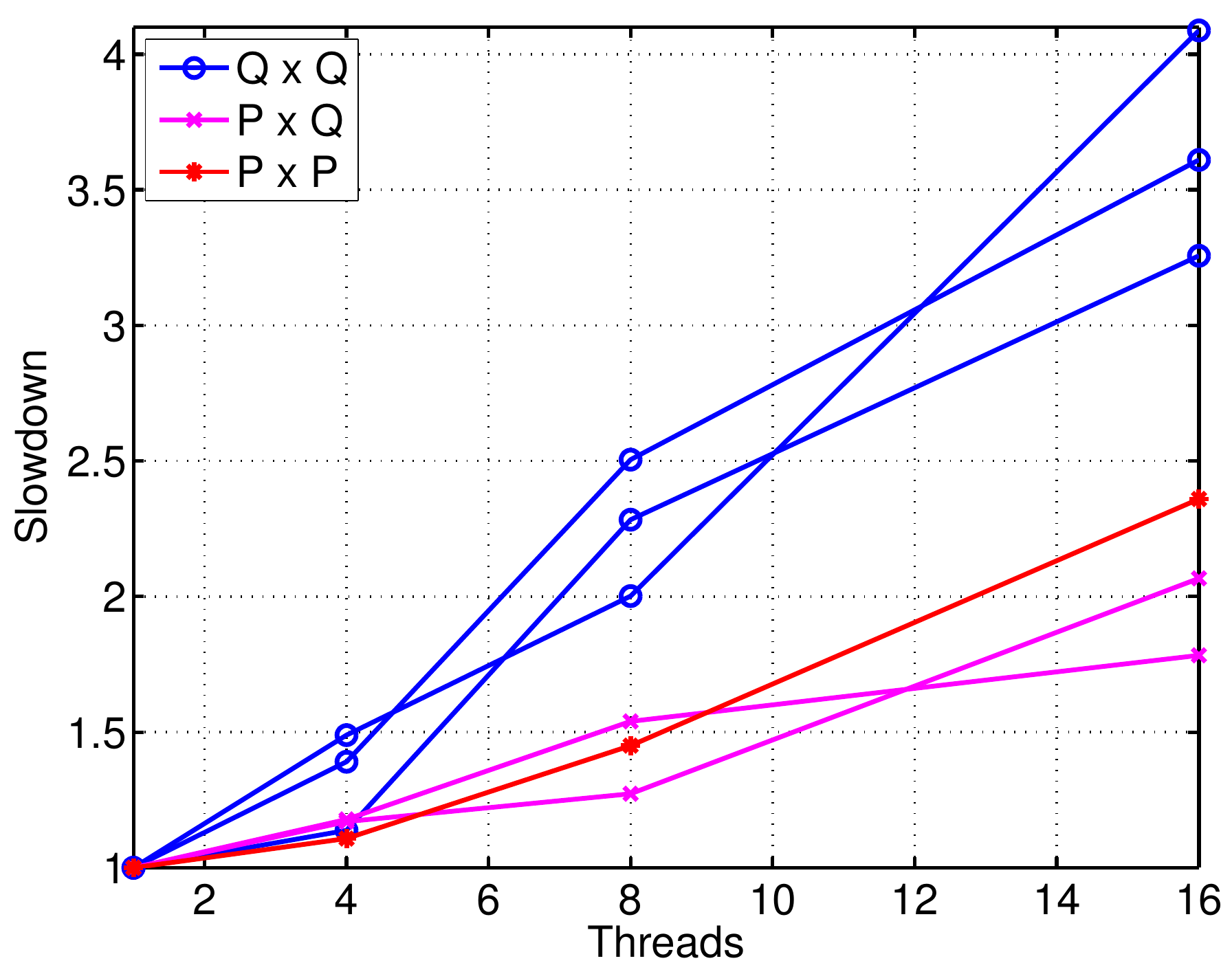}
\includegraphics[width=0.49\textwidth]{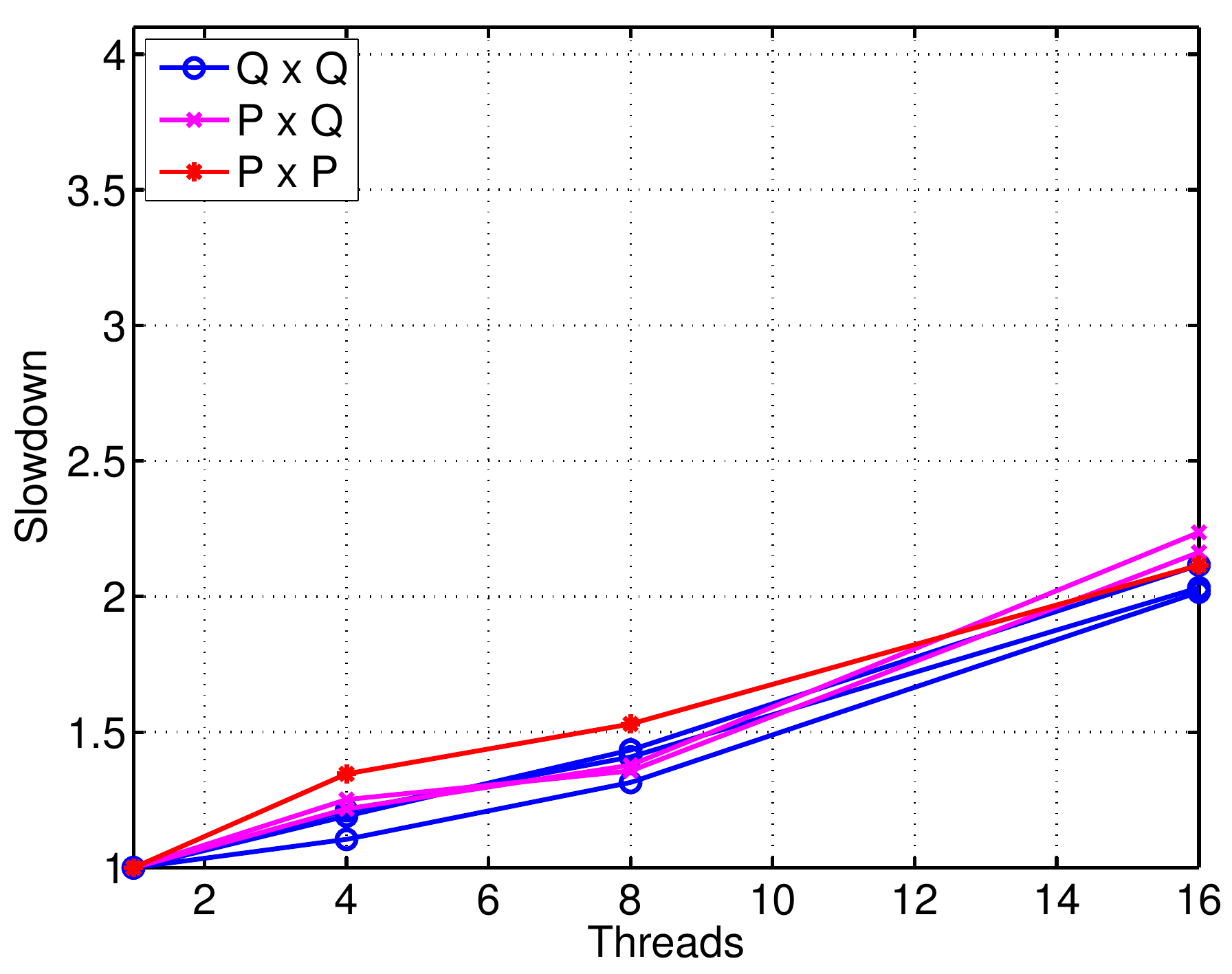}
\caption{Increase in individual task execution times for the complete execution for $P=50$, $Q=10$ (left) and for $P=300$, $Q=100$ (right).}
\label{fig:slowdownmix}
\end{figure}

To further understand the improved performance, we also look at the slow down profiles for the tasks. Figure~\ref{fig:slowdownmix} shows that the slowdown is significant, especially for the smallest tasks, for the execution trace where the tasks do not mix. However, for the mixed trace with larger tasks, the performance is perfect for all tasks. A slowdown factor of 2 would be expected due to the shared FPUs. This can be understood in the following way: The near field tasks performed well already from the start. The far field tasks did not perform as well, but this was for the case when they were executed in parallel on all threads. Here, there are fewer far field tasks running in parallel at each point in time, and the performance degradation is avoided.

Table~\ref{tab:nfff} shows the speedup and utilization results for the whole execution, confirming that the results are excellent if the task sizes are large enough.
\begin{table}
\caption{Performance results for the complete computation for two different parameter sets. $T_p$ is the execution time for $p$ threads, $S_p$ is the speedup, $S_p^*$ is the theoretical optimal speedup, and $U_p$ is the utilization.}
\label{tab:nfff}
\centering
\begin{tabular}{rrrrr}\hline
\multicolumn{5}{c}{$Q=10$, $P=50$}\\\hline
$p$ & $T_p$ [ms] & $S_p$ & $S_p/S_p^*$ & $U_p$\\\hline
1 &  244&  1.0  &  1.00 & 0.90\\
4 &  111&  2.2 &  1.10 & 0.55\\
8 &  137&  1.8&    0.44 &  0.29\\
16&  156&  1.6&    0.20 &  0.21\\\hline
\multicolumn{5}{c}{$Q=100$, $P=300$}\\\hline
1 & 1192 & 1  & 1.00 & 0.99 \\
4 &  401 & 3.0& 1.49 & 0.98\\
8 &  228 & 5.2& 1.31 & 0.98\\
16&  163 & 7.3& 0.92 & 0.96\\\hline
\end{tabular}
\end{table}







\section{Comparison with OpenMP}\label{sec:omp}
As explained in Section~\ref{sec:ompimp}, the same functionality can be achieved with an OpenMP implementation as with our SuperGlue implementation. The purpose of the experiments in this section is to evaluate the relative performance of the two implementations. The experiments are carried out both on a node of the Tintin cluster, described in the previous section, and on a local shared memory system with 4 sockets of Intel Xeon E5-4650 Sandy Bridge processors, yielding a total of 64 cores. On Tintin, the codes were compiled with gcc version 4.9.1 and OpenMP 4.0 was used, while on Sandy Bridge the compiler was gcc 6.3.0 combined with OpenMP 4.5. 

Experiments are first performed for the near and far field computations separately, and then for the full execution, except in one case where the far field data is missing\footnote{The data cannot be recreated as the Tintin system has been decommissioned.}. Results for the execution times for the SuperGlue implementation (SG) and the OpenMP implementation (OMP) are reported together with the gain $G_p$ in using SG compared with OMP on $p$ threads computed as
\begin{equation}
G_p=\frac{T^\mathrm{OMP}_p}{T^\mathrm{SG}_p}-1.
\end{equation}
Results for Tintin are shown in Tables~\ref{tab:sgomp1} and~\ref{tab:sgomp2} for the two cases with smaller and larger tasks, respectively.
\begin{table}[!htb]
\caption{Execution times in ms and the gain for SG compared with OMP for the problem with $Q=10$ and $P=50$ on the Tintin system.}
\label{tab:sgomp1}
\centering
\begin{tabular}{rrrrrrrrrr}\hline
& \multicolumn{3}{c}{Near field} & \multicolumn{3}{c}{Far field} &
\multicolumn{3}{c}{Combined}\\\hline
$p$ & SG & OMP & $G_p$ & SG & OMP & $G_p$ & SG & OMP & $G_p$\\\hline
1  & 222 & 229 &  3\% &  27 & 72 &   167\% & 244 & 285 & 17\%\\
4  &  66 &  86 & 30\% &  86 & 72 & $-$16\% & 111 & 134 & 21\%\\
8  &  42 &  50 & 19\% & 113 & 77 & $-$32\% & 137 & 110 &$-$20\%\\
16 &  36 &  41 & 14\% & 119 &143 &    20\% & 156 & 254 & 63\%\\\hline
\end{tabular}
\end{table}
\begin{table}[!htb]
\caption{Execution times in ms and the gain for SG compared with OMP for the problem with $Q=100$ and $P=300$ on the Tintin system.}
\label{tab:sgomp2}
\centering
\begin{tabular}{rrrrrrr}\hline
& \multicolumn{3}{c}{Near field} &
\multicolumn{3}{c}{Combined}\\\hline
$p$ & SG & OMP & $G_p$ & SG & OMP & $G_p$ \\\hline
1  & 1064 & 1065 &     0\% & 1192 & 1186 & $-$1\%\\
4  &  363 &  345 &  $-$5\% &  363 &  345 & $-$5\%\\
8  &  210 &  186 & $-$11\% &  210 &  186 & $-$11\%\\
16 &  145 &  139 &  $-$4\% &  145 &  139 & $-$4\%\\\hline
\end{tabular}
\end{table}
For the smaller task size, all near field computations are faster with SG, while results for the far field computations (that do not scale) are inconclusive. The overall computation is faster with SG for all cases, except 8 threads, with the largest gain for 16 threads.

The test case with larger task sizes is different in the sense that the near field computations become the dominating part of the computations. In this case the performance for the near field computations is slightly better for OpenMP, and carries over unchanged to the overall performance.

The results for the Sandy Bridge system are given in Tables~\ref{tab:sgomp3} and~\ref{tab:sgomp4}.
\begin{table}[!htb]
\caption{Execution times in ms and the gain for SG compared with OMP for the problem with $Q=10$ and $P=50$ on the Sandy Bridge system.}
\label{tab:sgomp3}
\centering
\begin{tabular}{rrrrrrrrrr}\hline
& \multicolumn{3}{c}{Near field} & \multicolumn{3}{c}{Far field} &
\multicolumn{3}{c}{Combined}\\\hline
$p$ & SG & OMP & $G_p$ & SG & OMP & $G_p$ & SG & OMP & $G_p$\\\hline
1  & 363 & 397 &  9\% &  78 & 80 &   3\% & 438 & 476 & 8\%\\
4  & 127 & 151 & 19\% &  91 &133 &  46\% & 166 & 260 & 56\%\\
8  &  69 &  79 & 14\% &  87 &151 &  74\% & 100 & 197 & 98\%\\
16 &  44 &  54 & 23\% &  89 &153 &  71\% & 107 & 170 & 60\%\\
32 &  29 &  83 &187\% & 102 &182 &  78\% & 135 & 237 & 75\%\\
64 &  34 & 106 &214\% & 122 &176 &  44\% & 141 & 535 &281\%\\
\hline
\end{tabular}
\end{table}
\begin{table}[!htb]
\caption{Execution times in ms and the gain for SG compared with OMP for the problem with $Q=100$ and $P=300$ on the Sandy Bridge system.}
\label{tab:sgomp4}
\centering
\begin{tabular}{rrrrrrrrrr}\hline
& \multicolumn{3}{c}{Near field} & \multicolumn{3}{c}{Far field} &
\multicolumn{3}{c}{Combined}\\\hline
$p$ & SG & OMP & $G_p$ & SG & OMP & $G_p$ & SG & OMP & $G_p$\\\hline
1  & 2069 & 2281 & 10\% &  250 & 276 &   10\% & 2318 & 2556 & 10\%\\
4  &  738 &  813 & 10\% &   89 & 100 &   12\% & 811 & 913 & 11\%\\
8  &  383 &  424 & 11\% &   50 &  64 &   29\% & 422 & 469 & 11\%\\
16 &  232 &  222 & $-$4\% & 33 &  45 &   34\% & 244 & 253 &  4\%\\
32 &  130 &  148 & 14\% &   32 &  64 &   98\% & 154 & 157 &  2\%\\
64 &  118 &  125 &  6\% &   43 &  70 &   65\% & 127 & 133 &  5\%\\
\hline
\end{tabular}
\end{table}
For the smaller task size, SG is faster in all cases, and the gain increases with the number of threads. For the larger task size, the gain for SG is largest for the far field computations, but again, the overall gain carries over from the near field computations. In this case, the gain does not increase with the number of threads.

The overall conclusions from the comparison is that SG is better at handling small task sizes, especially in combination with larger numbers of threads, while for large enough tasks, the performance of the two implementations is similar. 

\section{Summary}\label{sec:sum}
The most challenging aspect of the NESA algorithm from a task parallel point of view is not the hierarchical dependencies as might be expected, but rather the small task sizes. For sizes typical for a two-dimensional problem, the tasks are too small to provide scaling to more than a few threads. However, for sizes appropriate for a three-dimensional problem, the performance is close to optimal and a significant speedup is achieved. 

The task parallel programming model provided an easy way to implement a parallel code with very small changes to the original implementation. Moreover, the asynchronous nature of the task execution allowed for mixing of the different types of tasks, which proved to be the key to achieve high performance for the complete MVP including the hierarchical far field computation.

The conclusion is that it is possible to achieve excellent performance on shared memory systems for the NESA type of MVPs provided the task sizes can be made large enough, and that the task-based approach is promising for the development of a distributed three-dimensional solver.

The comparison with OpenMP shows that with some attention to the technical details, the same functionality can be obtained by OpenMP as with the SuperGlue task parallel framework. The performance of OpenMP is currently worse for small tasks, and larger numbers of threads, but may improve in future implementations of the standard.





\section*{Acknowledgments}
This article is based upon work from COST Action IC1406 High-Performance Modelling and Simulation for Big Data Applications (cHiPSET), supported by COST (European
Cooperation in Science and Technology).
The computations were performed on resources provided
by SNIC through Uppsala Multidisciplinary Center for Advanced Computational Science (UPPMAX) under Project p2009014.

\appendix
\section{Shared data types and tasks}\label{app:A}
Program~\ref{prg:type} shows the definition of the C++ class \texttt{SGMatrix}, which equips a \texttt{Matrix} with a handle such that it can be used as a protected shared data in a task parallel execution.
\begin{lstlisting}[numbers=left,numbersep=2pt, numberstyle=\tiny,
numberfirstline=true,language=C++,
caption={The SGMatrix data type, which protects a shared matrix with a handle.},
label=prg:type,mathescape]
#include "superglue.hpp"

extern SuperGlue *sgEngine;

class SGMatrix
{
  Handle<Options> *sg_handle;
  Matrix *M;
  bool trans;
public:
  SGMatrix(Matrix &m){
    sg_handle = new Handle<Options>;
    trans=false;
    M = &m;
  }
  // Further methods omitted here for brevity
};
\end{lstlisting}
%

Program~\ref{prg:task} shows the \texttt{SGTaskGemv} SuperGlue task class that provides the  MVP task.
\begin{lstlisting}[numbers=left,numbersep=2pt, numberstyle=\tiny,
numberfirstline=true,language=C++,
caption={The MVP task class.},
label=prg:task,mathescape]
class SGTaskGemv : public Task<Options,3>{
private:
  SGMatrix *A,*x,*y;
public:
  bool transA;
  enum{COL_MAJOR,ROW_MAJOR};
  SGTaskGemv(SGMatrix &A_,SGMatrix &x_,SGMatrix &y_)/*@\label{l:cs}@*/
  {  
    A = &A_;
    x = &x_;
    y = &y_;
    register_args();
  } /*@\label{l:ce}@*/
  void register_args(){
    Handle<Options> &hA = A->get_handle();
    Handle<Options> &hx = x->get_handle();
    Handle<Options> &hy = y->get_handle();
    register_access(ReadWriteAdd::read, hA);
    register_access(ReadWriteAdd::read, hx);
    register_access(ReadWriteAdd::add , hy);
    transA = false;
  }
  void run(){
    int M = A->get_matrix()->rows();
    int N = A->get_matrix()->cols();
    double *Mat= A->get_matrix()->get_data_memory();
    double *X  = x->get_matrix()->get_data_memory();
    double *Y  = y->get_matrix()->get_data_memory();
    cblas_dgemv(COL_MAJOR,transA,M,N, 1.0, Mat, M, 
                X, 1, 1.0, Y, 1);
  }
};
\end{lstlisting}
In the task constructor, lines \ref{l:cs}--\ref{l:ce}, the data is copied into the task, and the access type for each handle is registered. In the NESA algorithm, the input vector and the matrix are constant, so the registration of the read accesses for A and v could have been omitted. The commutative \textit{add} access type is used for the output vector to allow reordering of the accesses by different tasks. The alternative would be a \textit{write} access, which implies that tasks that access the same vector $y$ need to be executed in the same order as they are submitted, resulting in less flexibility and potential performance losses~\cite{TiLaBaMa15}. When a task is executed by the SuperGlue run-time system, the \texttt{run} method of the task is called.

\bibliographystyle{elsarticle-num}

\end{document}